\documentclass{rQUF2e}

\usepackage{epstopdf}
\usepackage{subfigure}

\usepackage{multirow,setspace,times,amssymb,amsmath,graphicx,color,rotating,subfigure,url}

\begin{document}

\title{Symmetric thermal optimal path and time-dependent lead-lag relationship: Novel statistical tests and application to UK and US real-estate and monetary policies}

\author{
HAO MENG$^\sharp$$\dag$$\ddag$,
HAI-CHUAN XU$^\sharp$$\dag$$\ddag$\thanks{$^\sharp$These authors contribute to the paper equally.},
WEI-XING ZHOU$^{\ast}$$\dag$$\ddag$\thanks{$^\ast$Corresponding author. Email:
  wxzhou@ecust.edu.cn or dsornette@ethz.ch}, and
DIDIER SORNETTE$^{\ast}$$\S$$\P$
\affil{$\dag$School of Business and Research Center for Econophysics,
  East Chine University of Science and Technology, Shanghai 200237,
  China\\
$\ddag$Department of Mathematics, School of Science, East China University of Science and Technology, Shanghai 200237, China\\
$\S$Department of Management, Technology and Economics, ETH Zurich, Scheuchzerstrasse 7, CH-8092 Zurich, Switzerland\\
$\P$Swiss Finance Institute, c/o University of Geneva, 40 blvd. Du Pont d¡¯Arve, CH 1211 Geneva 4, Switzerland}}

\maketitle

\begin{abstract}
  We present the symmetric thermal optimal path (TOPS) method to determine the time-dependent lead-lag relationship between two stochastic time series. This novel version of the previously introduced TOP method alleviates some inconsistencies by imposing that the lead-lag relationship should be invariant with respect to a time reversal of the time series after a change of sign. This means that, if `$X$ comes before $Y$', this transforms into `$Y$ comes before $X$' under a time reversal. We show that previously proposed bootstrap test lacks power and leads too often to a lack of rejection of the null that there is no lead-lag correlation when it is present. We introduce instead two novel tests. The first criterion, based on the free energy p-value $\rho$, quantifies the probability that a given lead-lag structure could be obtained from random time series with similar characteristics except for the lead-lag information. The second self-consistent test embodies the idea that, for the lead-lag path to be significant, synchronizing the two time series using the time varying lead-lag path should lead to a statistically significant correlation. We perform intensive synthetic tests to demonstrate their performance and limitations. Finally, we apply the TOPS  method with the two new tests to the time dependent lead-lag structures of house price and monetary policy of the United Kingdom (UK) and United States (US) from 1991 to 2011. We find that, for both countries, the TOPS paths indicate that interest rate changes were lagging behind house price index changes until the crisis in 2006-2007. The TOPS paths also suggest a catch up of the UK central bank and of the Federal Reserve still not being on top of the game during the crisis itself, as diagnosed by again the significant negative values of TOPS paths until 2008. Only later did the central banks interest rates as well as longer maturity rates lead the house price indices, confirming the occurrence of the transition to an era where the central bank is ``causally'' influencing the housing markets more than the reverse. The TOPS approach stresses the importance of accounting for change of regimes, so that similar pieces of information or policies may have drastically different impacts and developments, conditional on the economic, financial and geopolitical conditions. This study reinforces the view that the hypothesis of  statistical stationarity in economics is highly questionable.
\end{abstract}

{\textit{keywords:}} Lead-lag structure; Symmetric thermal optimal path; Statistical test; Housing market; Monetary policy
 \\




\section{Introduction}

The housing market, being a market of the largest single assets owned by households and firms, is strongly related to external financing options. Such options are usually in the form of mortgages, especially for private residential real estate, which are provided by banks or other financial institutions. Consequently, when the central bank lowers target interest rates, real estate financing options become more affordable. This in turn would lead to a shift in the demand for real-estate, which impacts its prices  \citep{Adams-Fuss-2010-JHE}. Furthermore, the housing market contributes to the well-being of the global economy by contributing to households' wealth and thus stimulating consumption  \citep{Tobin-1969-JMCB}.

In theory, the influence of monetary policy on the development of real estate prices is well described. Using its two main instruments, target interest rates and money supply, monetary authorities should be able to stimulate distressed economies or even disarm economic overheating. The target interest rate is therefore extremely important, because it directly determines the condition of money lending and liquidity in the financial market. A famous theoretical attempt of explaining the dependencies in the real estate market can be found in the so-called ``financial accelerator'' \citep{Bernanke-Gertler-Gilchrist-1994-RES}. According to this mechanism, rising house price leads to an increasing demand in credit volume so as to finance house purchases on an appreciated price level. Because most of these mortgages are also secured by the property itself, higher valuation improves the net worth of the household by raising its housing capital. Through this mechanism, households increase the capacity for borrowing, which in turn puts pressure on the development of real estate price. Finally, higher real estate price reduces the default risks of loans by increasing the value of the pledged collateral, which in turn allows the banks to enlarge their borrowing activities and ensures financial stability \citep{Daglish-2009-JBF,Koetter-Poghosyan-2010-JBF}.

The above-mentioned mechanisms underline the importance of real estate prices for bank lending and the credit market because they imply that boom-bust cycles can be reinforced and mutually transmitted between the credit market and the housing market, jeopardizing macroeconomic stability. Thus, monetary policy should be reactive to these kinds of fluctuations within the credit and housing market to ensure financial and long-run price stability.

However, empirical studies on the influence of central banks on such cyclical economic behavior of real estate prices and credit conditions are limited and controversial. \citet{Bordo-Jeanne-2002-IF} and \citet{Chen-Chou-Wu-2013-PER} find that the impact of interest rate movements on the development of asset prices triggered by the central banks has a highly non-linear effect depending on economic sentiment, which is mainly caused by the perception of market participant depending on the current phase of the cycle. Empirical evidence based on the vector auto-regression technique suggests that interest rate shocks have a significant effect on house prices \citep{Gupta-Jurglias-Kabundi-2010-EconM,Sa-Towbin-Wieladek-2014-JEEA}. Moreover, \citet{Dokko-Doyle-Kiley-Kim-Sherlund-Sim-Heuvel-2011-EconP} and \citet{Hume-Sentance-2009-JIMF} argue that the expansionary monetary policy, with its low levels of interest rates, has significantly contributed to the most recent house price boom. In contrast, there are also studies suggesting that housing market should have a larger effect on the growth of the economy, thus playing an important role in the monetary policy setting \citep{Lacoviello-2004-JHE,Attanasio-Leicester-Wakefield-2011-JEEA,Laibson-Mollerstrom-2010-EJ}.

Here, using an extension of the novel thermal optimal path technique \citep{Sornette-Zhou-2005-QF,Zhou-Sornette-2006-JMe,Zhou-Sornette-2007-PA} for the joint analysis of pairs of time series, we revisit the time-dependent lead-lag relationship between monetary policies and housing market of the United Kingdom and United States. The thermal optimal path method (abbr. TOP), proposed by \citet{Sornette-Zhou-2005-QF}, is situated in the time domain. But unlike other popular methods, it is a non-parametric approach. In some sense, it has similarities with the cross recurrence analysis \citep{Marwan-Kurths-2002-PLA}, which is also based on the construction of a ``distance matrix'' but extents this tool by the concept of a weighted average minimal path within the distance matrix. This allows one to better analyze non-linear correlations of time series based on a time-dependent lead-lag relationship. The quality of the empirical results obtained by the TOP method is promising, especially when comparing the method with other linear techniques. \citet{Sornette-Zhou-2005-QF} and \citet{Zhou-Sornette-2006-JMe,Zhou-Sornette-2007-PA} successfully apply it to investigate the dependence relationship between inflation and gross domestic product and inflation and unemployment in the USA, and obtain several new insights. \citet{Guo-Zhou-Cheng-Sornette-2011-PLoS1} extend the TOP method and applied it to investigate the lead-lag dependencies between the US stock market and monetary policies. \citet{Guo-Zhou-Cheng-2012-cnJMSC} apply the method to check the economy barometer effect of China stock market and no such effect was found.

Despite the major advantages of the TOP method, its use for routine empirical analyses still faces some difficulties. While the TOP method is efficient at identifying stable patterns between time series that have a well-established lead-lag structure, the impact of noise requires more investigation. Indeed, without prior knowledge, disentangling the genuine part of the signal from the contribution of noise is sometimes difficult. In order to assess the statistical significance of a lead-lag structure between two time series determined by TOP, \citet{Guo-Zhou-Cheng-Sornette-2011-PLoS1} propose a specifically designed bootstrapping test. However, as demonstrated in the following context, this bootstrap test has low power, in the sense that it fails to reject the null of no lead-lag structure in long-time series when the signal is actually present with not too large lead or lag in the cross-correlations. Another issue is that the identified TOP paths tend to be distorted or biased when dealing with weak lead-lag signals. We show below that this bias originates from the non-fully consistent weights attributed to the distances between the two time series along the time direction.

The paper is organized as follows. Section \ref{Sec:Method} summarizes the original TOP method, from which we develop the symmetric thermal optimal path method (abbr. TOPS) that remedies the above stated problems. Section \ref{Sec:StatTest} confirms the efficiency of the new TOPS method by way of illustration with several numerical experiments. We introduce two novel statistical tests specific to the TOPS method. Section \ref{Sec:Application} presents applications of these new proposed method to the identification of the time dependent lead-lag structures between house price and monetary policy of the United Kingdom and United States. Section \ref{Sec:Conclusion} summarizes.

\section{Method}
\label{Sec:Method}

\subsection{The thermal optimal path method (TOP)}

The TOP method has been proposed as a novel method to quantify the dynamical evolution of lead-lag structures between two time series \citep{Sornette-Zhou-2005-QF,Zhou-Sornette-2006-JMe,Zhou-Sornette-2007-PA}. Consider two standardized time series $X(t_X):t_X=0,\cdots,N-1$ and $Y(t_Y):t_Y=0,\cdots,N-1$, i.e., $X(t_X)$ and $Y(t_Y)$ have zero means and unit variance. We first form a distance matrix that allows us to compare systematically all values of the first time series $X(t_X)$ along the time axis $t_X$ with all the values of the second time series $Y(t_Y)$ along the time axis $t_Y$. As to the definition of distance, we can in principle use any definition that obeys the axioms to qualify as a genuine distance and that is sufficiently local to be attributed to each node. Here, we define the distance matrix $E_{X,Y}$ between $X$ and $Y$ \citep{Sornette-Zhou-2005-QF}, as the matrix of elements $\epsilon_-(t_X,t_Y)$ built by comparing the value of the first time series $X(t_X)$ at a given time $t_X$ with the value of the second time series $Y(t_Y)$
at another given time $t_Y$. Using the $L^2$ norm, this defines the scalar element $ \epsilon_{-}(t_X,t_Y)$ at row $t_X$ and column $t_Y$
\begin{equation}
  \epsilon_{-}(t_X,t_Y) = [X(t_X)-Y(t_Y)]^2.
  \label{Eq:DistMatrix}
\end{equation}
Scanning all possible values of $t_X$ and $t_Y$ allows us to construct the set of matrix elements $ \epsilon_{-}(t_X,t_Y)$ defining the distance matrix $E_{X,Y}$ between $X$ and $Y$. The value of $\epsilon_-(t_X,t_Y)$ measures the distance between the realizations of $X$ at time $t_X$ and $Y$ at time $t_Y$. Depending on the nature of the time series, it is interesting to use other distances, for instance \begin{equation}
\epsilon_{+}(t_X,t_Y)=[X(t_X)+Y(t_Y)]^2
\end{equation}
and
\begin{equation}
\epsilon_{\pm}(t_X,t_Y)=\min[\epsilon_{-}(t_X,t_Y), \epsilon_{+}(t_X,t_Y)]
\end{equation}
to deal with the anti-monotonic and non-monotonic relationship between two time series  \citep{Zhou-Sornette-2007-PA}. Our method introduced below applies the first distance without modifications.

The dependence relationship between the two time series is searched for in the form of a one-to-one mapping $t_Y=\phi(t_X)$ between the times ${t_X}$ of the first time series $X$ and the times ${t_Y}$ of the second time series such that the two time series are closest in a certain sense. Consider a simple example like $Y(t)=X(t-k)$ with $k>0$ is a positive constant. Then we get $\epsilon_-(t_X,t_Y)=0$ for $t_Y=t_X+k$. Detecting this dependence relationship then amounts to find the line with zero values which is parallel to the main diagonal of the distance matrix $E_{X,Y}$ with distance definition $\epsilon_-$. This line equivalently defines the affine mapping $t_Y=\phi(t_X)=t_X+k$.

A naive approach to determine the dependence relationship between two time series
could be thought to find the set of pairs $(t_X, t_Y)$ such that the elements of the distance matrix are
the smallest. This defines the mapping as
\begin{equation}
\phi(t_X)=\textrm{Min}_{t_Y}\{\epsilon(t_X,t_Y)\}~.
\label{Eq:LocalMinimization}
\end{equation}
This procedure analyzes each time $t_X$ independently, which causes two undesirable features. First, in the presence of noise, the function $t_Y=\phi(t_X)$ is quite discontinuous with numerous large jumps $\phi(t_X+1)-\phi(t_X)$ of amplitudes comparable to the total duration $N$ of the time series, at many times $t_X$. Second, with large probability, a given $t_X$ would be associated with several $t_Y$. Thus, there will be pairs of times $t_X<t_X'$ such that $\phi(t_X)>\phi(t_X')$, contradicting causality. Therefore, these two properties disqualify the simple method in Eq.~(\ref{Eq:LocalMinimization}) as a suitable construction of time dependence between two time series, since the lead-lag structure that would be obtained would be erratic, noisy and unreliable.

In general and for reasonable dependence structures, we expect the lead-lag relation between two time series to be a slowly varying function of time without unreasonable large jumps. To capture this insight, \citep{Sornette-Zhou-2005-QF} replace the mapping (\ref{Eq:LocalMinimization}) determined by a local minimization by a mapping obtained from the global minimization \begin{equation}
\textrm{Min}_{\{\phi(t_X), t_X=0,1,2,\ldots,N-1\}}~~~ E:= \sum_{t_X=0}^{N-1}|X(t_X)-Y(\phi(t_X))|
\label{Eq:GlobalMinimization}
\end{equation}
under the constraint of continuity expressed in discrete time
\begin{equation}
0 \leq \phi(t_X+1)-\phi(t_X) \leq 1~.
\label{Eq:ContinuousConstraint}
\end{equation}
In the continuous time limit of condition (\ref{Eq:ContinuousConstraint}), this amounts to imposing that the mapping $t_X \rightarrow t_Y = \phi(t_X)$ should be continuous. The continuity constraint makes the determination of $t_Y = \phi(t_X)$ a global optimization problem (\ref{Eq:GlobalMinimization}) rather than the simple local one (\ref{Eq:LocalMinimization}).

Actually, this type of problem has a long history and has been extensively studied under the name of the ``random directed polymer at zero temperature'' \citep{HalpinHealy-Zhang-1995-PR}. The distance matrix $E_{X,Y}$ can be interpreted as an energy landscape in the plane $(t_X,t_Y)$, in which the local distance $\epsilon(t_X,t_Y)$ is the energy associated with the node $(t_X,t_Y)$. The continuity constraint (\ref{Eq:ContinuousConstraint}) means that the mapping defines a path or ``polymer'' of $t_Y=\phi(t_X))$ with a ``surface tension'' to prevent discontinuities. The causality condition that $\phi(t_X)$ should be non-decreasing corresponds to the constraint that the polymer should be directed. Then we can translate the global minimization problem (\ref{Eq:GlobalMinimization}) into searching for the polymer configuration with minimum energy $E$ in the energy landscape. In the case where the two time series do have a dependence relationship, their corresponding energy landscape will exhibit coherent structures,  which can ``trap'' the polymer configuration and reveal the lead-lag structure.

\begin{figure}[!ht]
\centering
  \includegraphics[width=8cm]{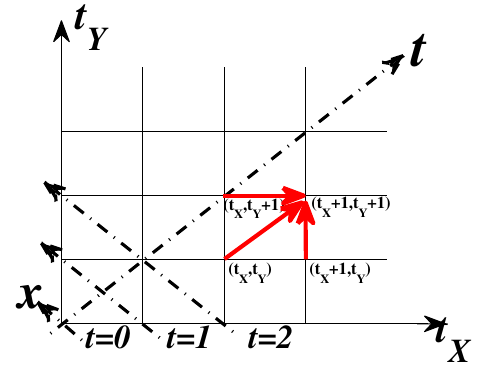}
  \caption{Schematic representation of the lattice $(t_X,t_Y)$ and of the rotated frame $(t,x)$. For clarity, we refer to the $(t_X,t_Y)$ coordinate system as the $\Box$ system and the $(t,x)$ coordinate system as the $\triangle$ system. The three arrows depict the three moves that are allowed from any node in one step, in accordance with the continuity and monotonicity constraint.}
\label{Fig:TOP Schematic Lattice}
\end{figure}

We use the transfer matrix method in the rotated coordinate system that allows one to solve this global optimization problem (\ref{Eq:GlobalMinimization}) in polynomial time \citep{Derrida-Vannimenus-Pomeau-1978-JPC,Derrida-Vannimenus-1983-PRB}. Fig.\ref{Fig:TOP Schematic Lattice} shows a schematic representation of the elementary steps underlying the transfer matrix method. The first (resp. second) time series is indexed by the time $t_X$ (resp. $t_Y$). A given node $(t_X,t_Y)$ in the $\Box$ lattice carries the ``potential energy'' or distance $\epsilon(t_X,t_Y)$. The $\triangle$ lattice is the rotated coordinate of $(t_X,t_Y)_{\Box} \rightarrow (t,x)_{\triangle}$. According to the continuity restriction (\ref{Eq:ContinuousConstraint}), the optimal path can either go horizontally by one step from $(t_X,t_Y)$ to $(t_X+1,t_Y)$, vertically by one step from $(t_X,t_Y)$ to $(t_X,t_Y+1)$ or along the diagonal from $(t_X,t_Y)$ to $(t_X+1,t_Y+1)$. Let us call $E(t_X,t_Y)$ the cumulative energy of the optimal path starting from some origin $(t_{1,0},t_{2,0})$ and ending at $(t_X,t_Y)$.
Then, the transfer matrix method is based on the following recursive relation:
\begin{equation}
  E(t_X,t_Y)=\epsilon(t_X,t_Y)+\textrm{Min}\{E(t_X-1,t_Y),E(t_X,t_Y-1),E(t_X-1,t_Y-1)\},
  \label{Eq:RecursiveEnergy}
\end{equation}
which means that the minimum energy path that reaches point $(t_X,t_Y)$ can only come from one of the three points $(t_X-1,t_Y)$, $(t_X,t_Y-1)$ and $(t_X-1,t_Y-1)$ preceding it. Then the minimum energy path reaching $(t_X,t_Y)$ is nothing but a recursive extension of the minimum energy path reaching one of these three preceding points. The global minimization procedure is fully determined once the starting and ending points of the paths are defined. To respect the fact that one of the two time series probably precede the other one, we allow the starting point to lie anywhere on the horizontal axis $t_Y=0$ or on the vertical axis $t_X=0$. Similarly, the ending point could lie anywhere on the horizontal axis $t_Y=N-1$ or on the vertical axis $t_X=N-1$. The minimum energy path over all possible starting and ending points is the solution of the global optimization (\ref{Eq:GlobalMinimization}) under the constraint (\ref{Eq:ContinuousConstraint}), whose equation $t_X \rightarrow t_Y=\phi(t_X)$ defines the dependence relationship between the two time series.

In this optimization program, it is assumed that the energy landscape $E_{X,Y}$ derived from the time series $X$ to $Y$ is made only of useful information. In reality, the two time series are noisy. Their corresponding energy landscape can be expected to contain spurious patterns, which may hinder the determination of the global optimal path, leading to incorrect inferred lead-lag relationships. In the case where the noise dominates the energy landscape, small changes in the distance matrix may lead to very large jumps in the optimal path \citep{HalpinHealy-Zhang-1995-PR,Jogi-Sornette-1998-PRE}. To obtain a robust lead-lag path, \citet{Sornette-Zhou-2005-QF} propose to introduce ``thermal excitations'' around the path, so that path configurations with slightly larger global energies are allowed to contribute to  the conformation of the optimal path with probabilities proportional to their Boltzmann weights. They specify the probability of a given path configuration with energy $\Delta E$ above the absolute minimum energy path by a Boltzmann-Gibbs factor $\exp(-\Delta E / T)$, where the ``temperature'' $T$ quantifies how much deviation from the minimum energy is allowed. For $T \rightarrow 0$, the probability for selecting a path configuration with incremental energy $\Delta E$ above the absolute minimum energy path goes to zero. Increasing $T$ allows one to sample more and more paths around the minimum energy path, thus giving an average ``optimal thermal path'' over a larger and larger number of path conformations. This tends to wash out possible idiosyncratic dependencies of the path conformation on the specific realization of the noises decorating the two time series. The introduction of a finite temperature and the thermal averaging over path weighted by the Boltzmann-Gibbs factor can be interpreted as a renormalization of the energy (or distance) landscape in which the path configuration are sampled, decimating the details of the local distance matrix to obtain more robust and smooth paths. Of course, for too large temperatures, one would lose the information of the lead-lag relationships between two time series since it samples too many paths independent of the set of local distances between the realizations of the two time series. There is therefore a compromise between not extracting too much from the spurious noise and washing out too much of the relevant signal.

It is convenient to use the $\triangle$ frame $(t,x)$, as defined in Fig.~\ref{Fig:TOP Schematic Lattice}, to implement these ideas. As illustrated in Fig.~\ref{Fig:TOP Schematic Lattice}, the transformation from the coordinates $(t_X,t_Y)_{\Box}$ to $(t,x)_{\triangle}$ is
\begin{equation}
\left\{
  \begin{array}{ccl}
  t &=& t_Y+t_X\\
  x &=& t_Y-t_X
  \end{array}
\right.
\label{Eq:AxesTransform}
\end{equation}
The optimal path for two synchronized time series is the main diagonal, so that deviations from the diagonal quantify the lead or lag times between the two time series. It is thus convenient to use the $\triangle$ frame $(t,x)$ in which the second coordinate $x$ quantifies the deviation from the main diagonal, hence the lead or lag time between the two time series. With the constraint that the path is directed, we can interpret $t$ as an effective time and $x$ as the position of a path at that ``time'' $t$. Then, the optimal thermal averaged path  trajectory $\langle x(t) \rangle$ is given by
\begin{equation}
\langle x(t) \rangle = \sum_{x} xG_{\triangle}(t,x)/G_{\triangle}(t).
\label{Eq:ThermalAveragePath}
\end{equation}
where $G_{\triangle}(t,x)$ is the sum of Boltzmann factors over all paths emanating from $(0,0)_{\triangle}$ and ending at $(t,x)_{\triangle}$ and $G_{\triangle}(t) = \sum_{x} G_{\triangle}(t,x)$. As illustrated in Fig.~\ref{Fig:TOP Schematic Lattice} and Eq.~(\ref{Eq:RecursiveEnergy}), in the $\Box$ frame, to arrive at $(t_X+1,t_Y+1)$, the path can come from $(t_X+1,t_Y)$ vertically, $(t_X,t_Y+1)$ horizontally, or $(t_X,t_Y)$ diagonally. The recursive equation on the Boltzmann weight factor is thus
\begin{equation}
G(t_X+1,t_Y+1)=[G(t_X+1,t_Y)+G(t_X,t_Y+1)+G(t_X,t_Y)]e^{-\epsilon(t_X+1,t_Y+1)/T},
\label{Eq:Recursive_GSquare}
\end{equation}
where $\epsilon(t_X+1,t_Y+1)$ is the local energy determined by the distance matrix element at node $(t_X+1,t_Y+1)$. In term of the coordinate transformation (\ref{Eq:AxesTransform}), it can be rewritten in the following form
\begin{equation}
G_{\triangle}(t+1,x)=[G_{\triangle}(t,x-1)+G_{\triangle}(t,x+1)+G_{\triangle}(t-1,x)]e^{-\epsilon(t+1,x)/T}.
\label{Eq:Recursive_GTriangle}
\end{equation}
In statistical physics, $G_{\triangle}(t,x)$ is called the partition function constrained to $x$ while $G_{\triangle}(t)$ is the total partition function at $t$. Naturally, $G_{\triangle}(t,x)/G_{\triangle}(t)$ is nothing but the probability for a path to be at position $x$ at time $t$, which is a compromise between minimizing the energy and maximizing the combinatorial weight of the number of paths with similar energy in a neighborhood (similar to an ``entropy''). Eq.~(\ref{Eq:ThermalAveragePath}) indeed defines $\langle x(t) \rangle$ as the thermal average of the local time lag at $t$ over all possible lead-lag configurations suitably weighted according to the Boltzmann factor. This justifies to call it the ``thermal averaged path''.

As we discussed in the preceding text, once we get an origin and an ending point, for instance $(t_0,x_0)_{\triangle}$ with position $x_0$ as starting point and $(t_n,x_n)$ with position $x_n$ as ending point, we can determine the thermal optimal path $\langle x(t) \rangle$ under a certain temperature $T$. We can also define a cost ``energy'' $e_T$ for this path as the thermal average of the distance $\epsilon(t,x)_{\triangle}$ between the two time series:
\begin{equation}
e_T=\frac{1}{2N-|x_0|-|x_N|-1}\sum^{2N-|x_N|-1}_{t=|x_0|}\sum_{x}\epsilon(t,x)G_{\triangle}(t,x)/G_{\triangle}(t),
\label{Eq:FreeEnergy}
\end{equation}
where $N$ is the length of the time series and $|x_0|$ and $|x_n|$ are the absolute values of the positions of the starting and ending points.

\subsection{Thermal optimal path method with time-reversed symmetric weight (TOPS)}

The thermal optimal path method (TOP) starts with the allocation of weights to each node. Each node $(t_X,t_Y)$ is like a city, with energy $\epsilon(t_X,t_Y)$ as its jetton (the smaller the better), which competes with others (nodes) to welcome the final optimal paths to pass through. This ``optimal'' path is determined as the best compromise for the route over different competing nodes.

By construction, the weight of a given node is independent of whether we study the lead-lag dependence from left-to-right or from right-to-left. Consider two time series such that $X(t_X)$ leads $Y(t_Y)$ in a representation where the times $t_X$ and $t_Y$ are flowing from past to present. Then, reversing the time arrows, we can state equivalently that $Y(t_Y)$ precedes $X(t_X)$ when studying the lead-lag structures from future to past. Technically, this implies that the probability weight for a path to be present on a given node is independent on whether the path is determined recursively from left-to-right (past-to-present) or from right-to-left (future-to-past).

\definecolor{myred}{rgb}{1,0.4,0}
\definecolor{myyellow}{rgb}{1,0.8571,0.1429}
\definecolor{mygreen}{rgb}{0,0.5,0.4}
\begin{figure}[!ht]
\centering
  \includegraphics[width=7cm]{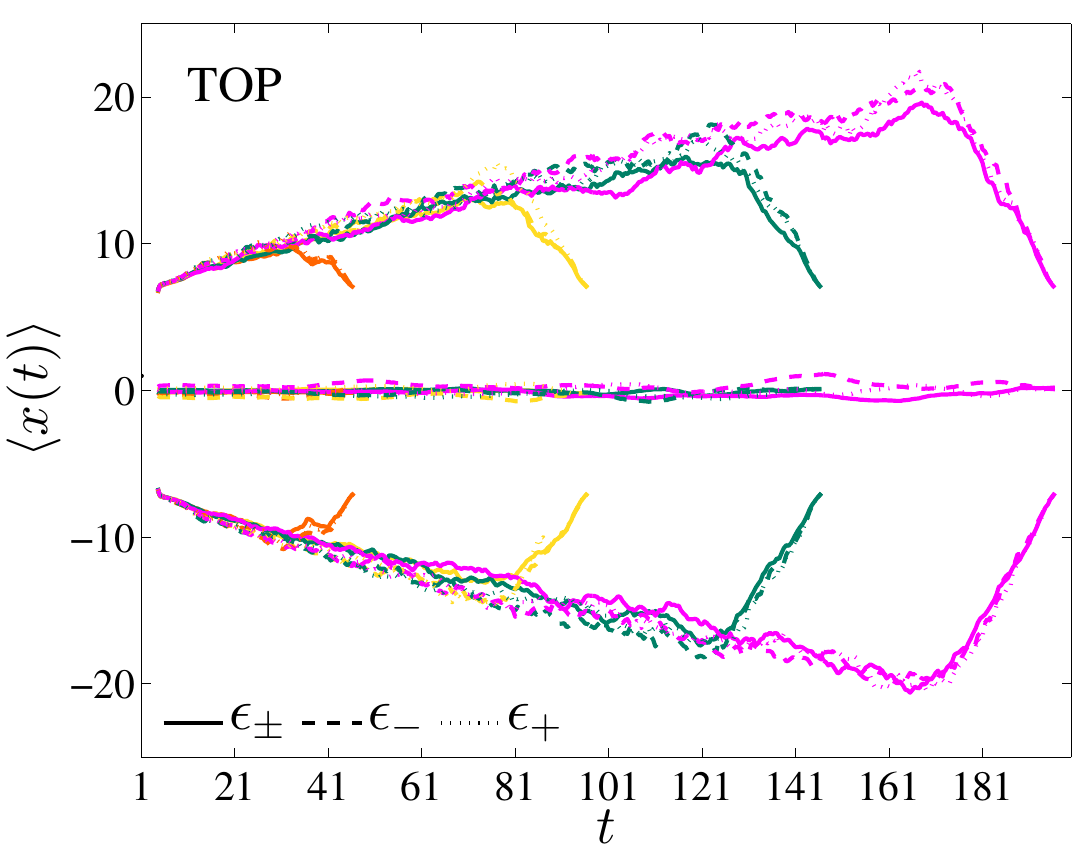}
  \includegraphics[width=7cm]{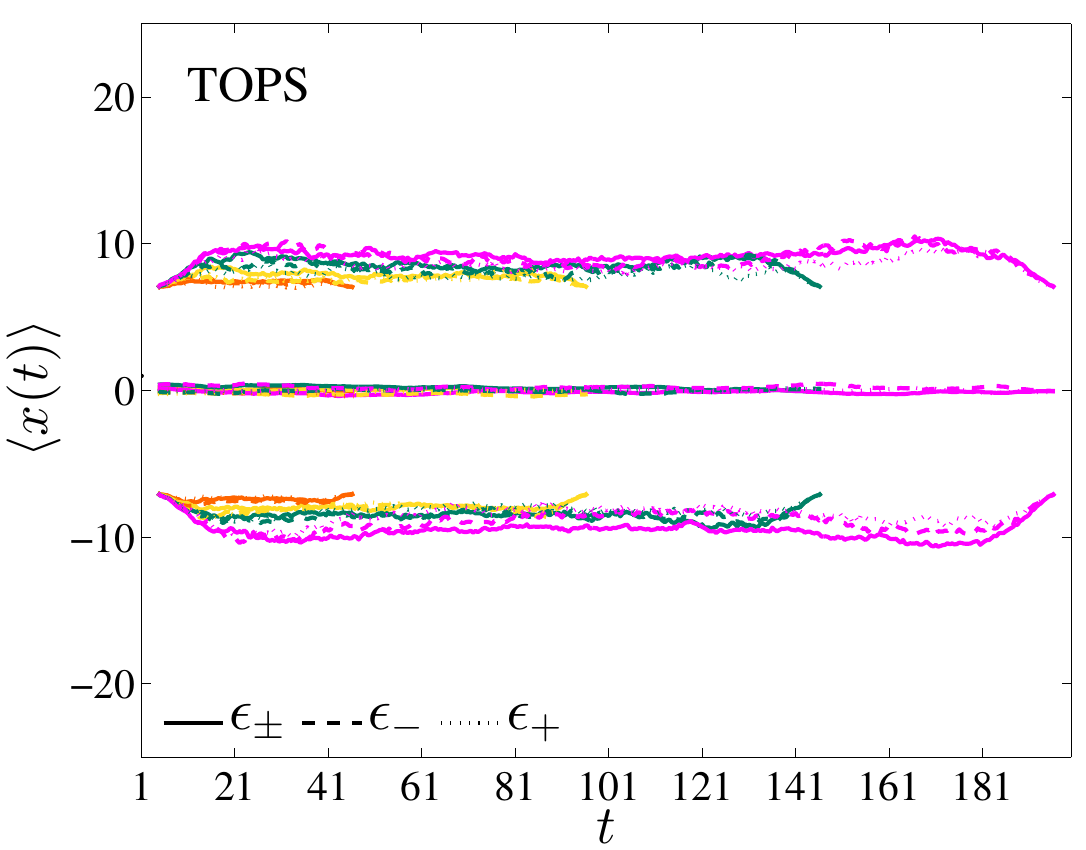}
  \caption{(color online) 95\%, 5\% quantiles and the averages over a population of 1000 random distance matrix configurations of four experiment groups. The distance matrices of each group are constructed using random time series, which are generated from a uniform distribution, with length $N=50$ (\textcolor{myred}{---}), 100 (\textcolor{myyellow}{---}), 150 (\textcolor{mygreen}{---}), 200 (\textcolor{magenta}{---}) respectively. Different line styles encode the different distances used. For the left panel (TOP), the weights of the paths at each node are determined by the recursive process (\ref{Eq:Recursive_GTriangle}). For the right panel (TOPS), the weights of the paths at each node are determined by a time-reversal invariant weight process presented in the text.}
  \label{Fig:BS Curve Length TOP/TOPS}
\end{figure}

In the case of random distance matrices for which there is no preferred lead-lag structure, the above  symmetry argument implies that the ensemble of random optimal paths should be symmetrically distributed around the zero-lag line $x(t)=0$ and should also be symmetric with respect to an inversion of the time axis, reflecting the absence of any preferred time arrow in the random configuration. But the left panel of Fig.~\ref{Fig:BS Curve Length TOP/TOPS} shows that the $5\%$ and $95\%$ quantiles over a population of $1000$ random optimal path configurations obtained over $1000$ realizations of random distances, whose nodes are weighted by the left-to-right recursive process (\ref{Eq:Recursive_GTriangle}), do not obey the time-reversed invariance symmetry. This breaking of the time-reversed symmetry makes it problematic to interpret confidence levels of optimal paths in the presence of noise, since it would lead to much larger confidence bounds (and thus less discriminating power), the further to the right (the larger along the time axis) we analyze the lead-lag structure.

\definecolor{mybrown}{rgb}{0.87,0.49,0}
\begin{figure}[!ht]
\centering
  \includegraphics[width=8cm]{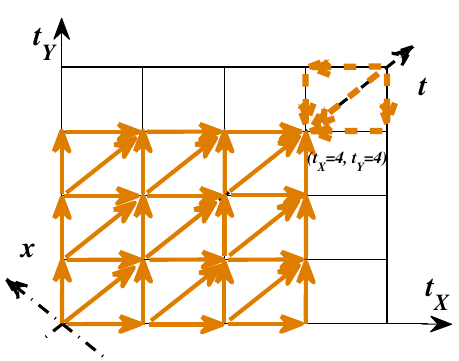}
  \caption{(color online) A simple schematic representation of the thermal optimal path method with time-reversed symmetric weight. The energy landscape is constructed by two time series of length $N=5$ ($t_X,t_Y=1,2,3,4,5$). The \textcolor{mybrown}{$\rightarrow$}'s denote that the weight process is recursively from left-to-right (time-forward direction) and the \textcolor{mybrown}{$\dashrightarrow$}'s indicate
  that the weight process is recursively from right-to-left (time-reversed direction).}
  \label{Fig:TOPS Schematic Lattice}
\end{figure}

The origin of this breaking of time-reversed symmetry lies in the different numbers of paths contributing to the probability weight attributed to a given node when calculated from left-to-right compared from the construction of the probability weight on that same node from right-to-left. As illustrated in Fig.~\ref{Fig:TOPS Schematic Lattice}, the weight of node $(t_X=4,t_Y=4)$ is recursively contributed by $15$ preceding nodes when searching from left-to-right (time-forward direction). But only $3$ nodes contribute to its weight when searching from right-to-left (time-backward direction). To remedy this problem, we propose to construct a time-reversed invariant node weight process, obtained by taken the average of the weights determined along the time-forward and time-backward directions. In other words, we modify the optimal thermal path trajectory calculated according to expression (\ref{Eq:ThermalAveragePath}) into
\begin{equation}
\langle x(t) \rangle = \sum_{x}x\frac{\overrightarrow{G_{\triangle}}(t,x)/\overrightarrow{G_{\triangle}}(t) + \overleftarrow{G_{\triangle}}(t,x)/\overleftarrow{G_{\triangle}}(t)}{2}~,
\label{Eq:ThermalAveragePath TOPS}
\end{equation}
where $\overrightarrow{G_{\triangle}}(t,x)$ and $\overrightarrow{G_{\triangle}}(t)$ are identical to $G_{\triangle}(t,x)$ and $G_{\triangle}(t)$ in Eq. (\ref{Eq:Recursive_GTriangle}), and where the arrow $\rightarrow$ denotes that the recursive weight process is along the time-forward direction. In contrast, $\overleftarrow{G_{\triangle}}(t,x)$ is the sum of Boltzmann factors over all paths emanating from an ending point, for instance $(N-1,N-1)_{\Box}$, and ending at the same node $(t,x)_{\triangle}$ as $\overrightarrow{G_{\triangle}}(t,x)$. It is nothing but the partition function constrained to the same $x$ as $\overrightarrow{G_{\triangle}}(t,x)$ when determining the weights from right-to-left and the arrow $\leftarrow$ denotes that the recursive weight process is along the time-backward direction. Similarly, $\overleftarrow{G_{\triangle}}(t) = \sum_{x} \overleftarrow{G_{\triangle}}(t,x)$. As shown in the right panel of Fig.~\ref{Fig:BS Curve Length TOP/TOPS}, by using the recursive weight (\ref{Eq:ThermalAveragePath TOPS}) instead of (\ref{Eq:ThermalAveragePath}), the widening trend is essentially eliminated and the ensemble of random paths obeys time-reversed invariance. In a similar way, we modify the cost energy $e_T$ of equation (\ref{Eq:FreeEnergy}) into
\begin{equation}
e_T=\frac{1}{2N-|x_0|-|x_N|-1}\sum^{2N-|x_0|-|x_n|}_{t=x_0}\sum_{x}\epsilon(t,x)\frac{\overrightarrow{G_{\triangle}}(t,x)/\overrightarrow{G_{\triangle}}(t) + \overleftarrow{G_{\triangle}}(t,x)/\overleftarrow{G_{\triangle}}(t)}{2}
\label{Eq:FreeEnergy TOPS}
\end{equation}

\section{Numerical experiments and test of statistical significance}
\label{Sec:StatTest}

\subsection{Comparison between TOP and TOPS on two synthetic time series with varying lead-lag structure}

We now present synthetic tests of the efficiency of the thermal optimal path method with the time-reversed symmetric weight process (abbr. TOPS) to detect multiple changes of regime in the lead-lag structure between two time series. Let us consider two stationary time series $X(t_X)$ and $Y(t_Y)$ constructed as follows:
\begin{equation}
Y(t_Y) =
\left\{
   \begin{array}{lrl}
    aX(t_Y-30)+\eta, &  1 &\leq t_Y \leq 100\\
    aX(t_Y-15)+\eta, &100 &< t_Y \leq 200\\
    aX(t_Y)+\eta,    &200 &< t_Y \leq 300\\
    aX(t_Y+15)+\eta, &300 &< t_Y \leq 400\\
    aX(t_Y+30)+\eta, &400 &< t_Y \leq 500
   \end{array}
\right.
\label{Eq:Synthetic Lead-Lag Model}
\end{equation}
where $a$ is a constant and the noise $\eta \sim N(0,\sigma_{\eta})$ is serially uncorrelated. $X(t)$ itself is generated by the first-order auto-regressive process
\begin{equation}
X(t_X)=bX(t_X-1)+\xi,
\label{Eq:FirstOrder Auto-regressive}
\end{equation}
where $b < 1$ and the noise $\xi \sim N(0,\sigma_{\xi})$ is serially uncorrelated. By construction, the time series $Y$ is lagging behind $X$ with lag $\tau=30,15$ and $0$ in three successive time periods of $100$ time steps each. Then, $Y$ leads $X$ with a leading time $-\tau=15$ and $30$ in the last two periods of $100$ time steps each. The factor $f=\sigma_{\eta}/\sigma_{\xi}$ quantifies the amount of noise degrading the dependence relationship between $X(t_X)$ and $Y(t_Y)$. Small $f$ values lead to a strong dependence relationship and large $f$'s correspond to the situation where $Y(t_Y)$ is mostly noise and is weakly related to $X(t_X)$ (and become unrelated in the limit $f \rightarrow \infty$). Here, we take $a=0.8$, $b=0.7$ and $f=0.2$.

\begin{figure}[!ht]
\centering
  \includegraphics[width=7cm]{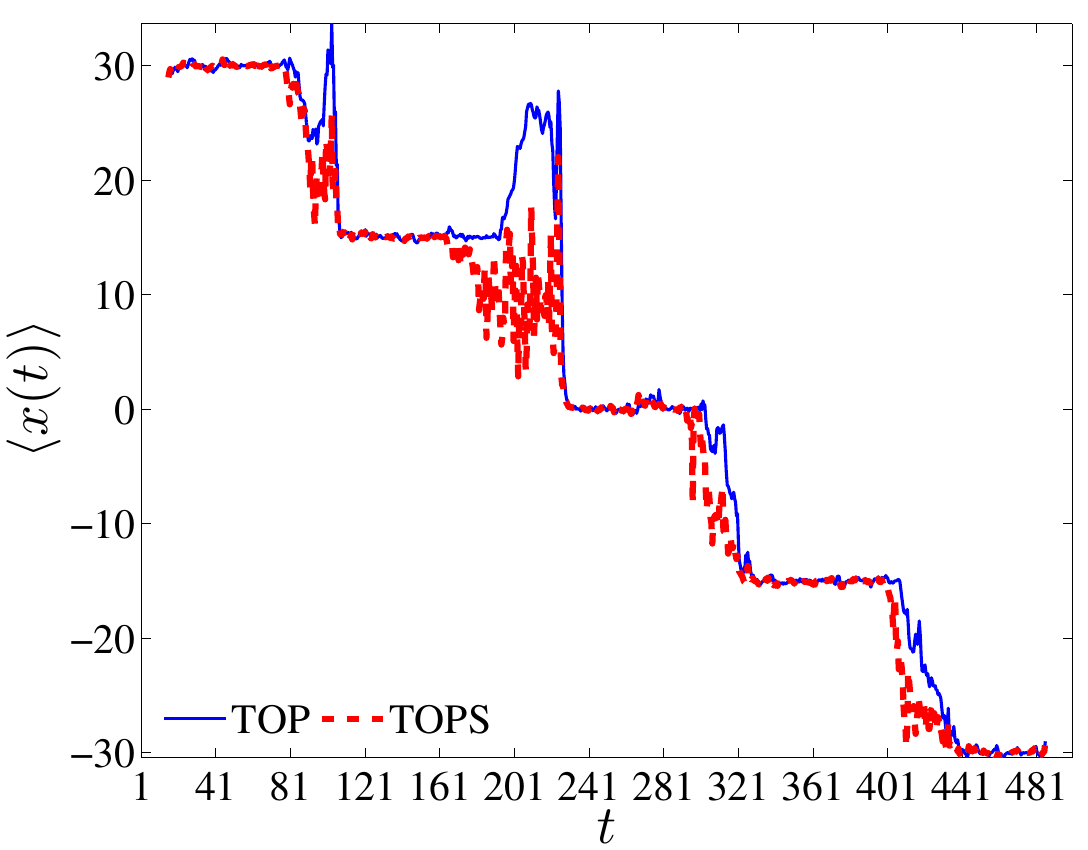}
  \caption{(color online) Thermal optimal path $\langle x(t) \rangle$ recovered by applying the TOP and TOPS methods to the synthetic time series generated according to expression (\ref{Eq:Synthetic Lead-Lag Model}). The temperature used here is $T=2$ and the chosen distance is $\epsilon_{-}(t_X,t_Y)$ defined in Eq.~(\ref{Eq:DistMatrix}).}
  \label{Fig:TOPS TOP Path AR}
\end{figure}

We determine the average thermal paths $\langle x(t) \rangle$ using both the original TOP method and the TOPS method. As discussed above, we take $61$ starting points and $61$ ending points into account, which are $(t_X=0,t_Y=0)$, $(t_X=0,t_Y=i)$, $(t_X=i,t_Y=0)$, $(t_X=N,t_Y=N)$, $(t_X=N,t_Y=N-i)$ and $(t_X=N-i,t_Y=N)$ for $i=1,2,\cdots,30$, where $N=500$ is the length of time series. The path with the minimum average energy defined by expression (\ref{Eq:FreeEnergy}) and (\ref{Eq:FreeEnergy TOPS}) respectively over these $61 \times 61$ paths is the thermal optimal path we are looking for in each scheme (TOP and TOPS). Fig.~\ref{Fig:TOPS TOP Path AR} depicts the thermal optimal path of the synthetic lead-lag structures constructed from (\ref{Eq:Synthetic Lead-Lag Model}) obtained by using the TOP and TOPS methods respectively. The time lags in the five time periods are recovered clearly by both TOP and TOPS. This numerical experiment is a first indication that the modification on the weight process does not impair the power of the method to detect complex varying lead-lag patterns. The comparison of the two paths even suggests a slight improvement of TOPS over TOP, in the sense that the TOPS lead-lag path exhibits smaller biases and less delay in capturing the changes of lead-lag.

\subsection{Bootstrapping curves test}

In order to test the statistical significance of an obtained thermal optimal path of the lead-lag structure between two time series, \citet{Guo-Zhou-Cheng-Sornette-2011-PLoS1} propose a bootstrapping test based on the idea of comparing it with the paths obtained from random time series. But in order for the comparison to be meaningful, the idea is to use random structures closely related to the initial one, except for the absence of any lead-lag structure. \citet{Guo-Zhou-Cheng-Sornette-2011-PLoS1} adopt the following procedure, which is specific to the thermal optimal path method. Given the two time series $\{X(t_X)\}$ and $\{Y(t_Y)\}$ to analyze, the idea is to compare the thermal optimal path $\langle x(t) \rangle$ obtained with the TOP/TOPS analysis at a given temperature $T$ to those obtained using time series pairs transformed by random reshuffling of the sequences of $\{X(t_X)\}$ and $\{Y(t_Y)\}$. Reshuffling $n$ times provides $n$ paths $x_i(t)$ associated with the $n$ random energy landscapes for $i=1,2,\cdots,n$. For each $t$, out of the $n$ reshuffled time series, we determine the $5\%$ and $95\%$ quantiles denoted in $x_{5\%}$ and $x_{95\%}$, shown in Fig.~\ref{Fig:BS Curve Length TOP/TOPS}. The bootstrapping test concludes that the thermal optimal path $\langle x(t) \rangle$ at time $t$ is statistically different from zero at the significance level of $95\%$, if $\langle x(t) \rangle$ is smaller that $x_{5\%}$ or larger than $x_{95\%}$.

A nice property of this bootstrapping test is that it can detect the statistical significance of the optimal path locally. Plotting the thermal optimal path $\langle x(t) \rangle$ together with the $x_{5\%}$ and $x_{95\%}$ confidence bounds on the same plot indicates clearly at which times $t$ the $\langle x(t) \rangle$ values are significantly outside the $90\%$ confidence interval.

However, the power of this bootstrapping test is low, in the sense that it has a low
probability to correctly rejects the null hypothesis when the null hypothesis is false. Here, the
null is the absence of non-zero lead-lag. The bootstrapping test
of \citet{Guo-Zhou-Cheng-Sornette-2011-PLoS1} unfortunately
concludes much too often on the absence of a non-zero lead-lag when there is truly one.
This can be understood intuitively from the way the bootstrapping test is
constructed: the closer the real lead-lag to $x(t)=0$, the larger the chance that
it lies within the $x_{5\%}$ and $x_{95\%}$ confidence bounds.
A small or zero lead-lag value $x(t) \approx 0$ is likely to be
interpreted as being a diagnostic of the absence of inter-dependence because the
bootstrapping test will conclude that the lead-lag path is undistinguishable
from those obtained from random time series. Of course,
this is incorrect since two time series can be strongly contemporaneously correlated.

Another problem is that the power of the bootstrapping test of \citet{Guo-Zhou-Cheng-Sornette-2011-PLoS1} weakens when the time series are longer. This results from the widening of the $x_{5\%} \sim x_{95\%}$ band with the length of the time series. As shown in Fig.~\ref{Fig:BS Curve Length TOP/TOPS}, time series with longer durations exhibit wider $x_{5\%} \sim x_{95\%}$ bands, even though we have removed the time trend (right panel) with the symmetric TOPS method. This is due to the widening of the random excursions of random paths in random energy landscapes as their length increases. This undesirable feature of the $x_{5\%} \sim x_{95\%}$ band dramatically impairs the power of the bootstrapping test. To quantify this statement, consider the three pairs of time series with the following different synthetic lead-lag structures:
\begin{subequations}
\begin{equation}
Y(i) =
\left\{
   \begin{array}{lrl}
    aX(i-30)+\eta,~~~~~&1   &\leq i \leq 100\\
    aX(i-15)+\eta,~~~~~&100 &< i \leq 200\\
    aX(i)+\eta,   ~~~~~&200 &< i \leq 300\\
    aX(i+15)+\eta,~~~~~&300 &< i \leq 400\\
    aX(i+30)+\eta,~~~~~&400 &< i \leq 500
   \end{array}
\right.
\label{subEq:Synthetic A}
\end{equation}
\vspace{-5mm}
\begin{equation}
Y(i) =
\left\{
   \begin{array}{lrl}
    aX(i-30)+\eta,  ~~~~~&  1 &\leq i \leq 200\\
    aX(i-15)+\eta,  &200 &< i \leq 400\\
    aX(i)+\eta,     &400 &< i \leq 600\\
    aX(i+15)+\eta,  &600 &< i \leq 800\\
    aX(i+30)+\eta,  &800 &< i \leq 1000
   \end{array}
\right.
\label{subEq:Synthetic B}
\end{equation}
\vspace{-5mm}
\begin{equation}
Y(i) =
\left\{
   \begin{array}{lrl}
    aX(i-60)+\eta,~~~~~&1   &\leq i \leq 200\\
    aX(i-30)+\eta,~~~~~&200 &< i \leq 400\\
    aX(i)+\eta,   ~~~~~&400 &< i \leq 600\\
    aX(i+30)+\eta,~~~~~&600 &< i \leq 800\\
    aX(i+60)+\eta,~~~~~&800 &< i \leq 1000
   \end{array}
\right.
\label{subEq:Synthetic C}
\end{equation}
\end{subequations}
Here, we still take $a=0.8$, $b=0.7$ and $f=0.2$ so that they have strong lead-lag signals. By construction, Eq.~(\ref{subEq:Synthetic B}) leads to the same lead-lag time values as Eq.~(\ref{subEq:Synthetic A}), but with double length of the time series $X(t_X)$ and $Y(t_Y)$. Comparing with Eq.~(\ref{subEq:Synthetic A}), Eq.~(\ref{subEq:Synthetic C}) doubles both the time lag values and the length of the time series.

\begin{figure}[!ht]
\centering
  \raisebox{3cm}{\textsf{A}}
  \includegraphics[width=4.5cm]{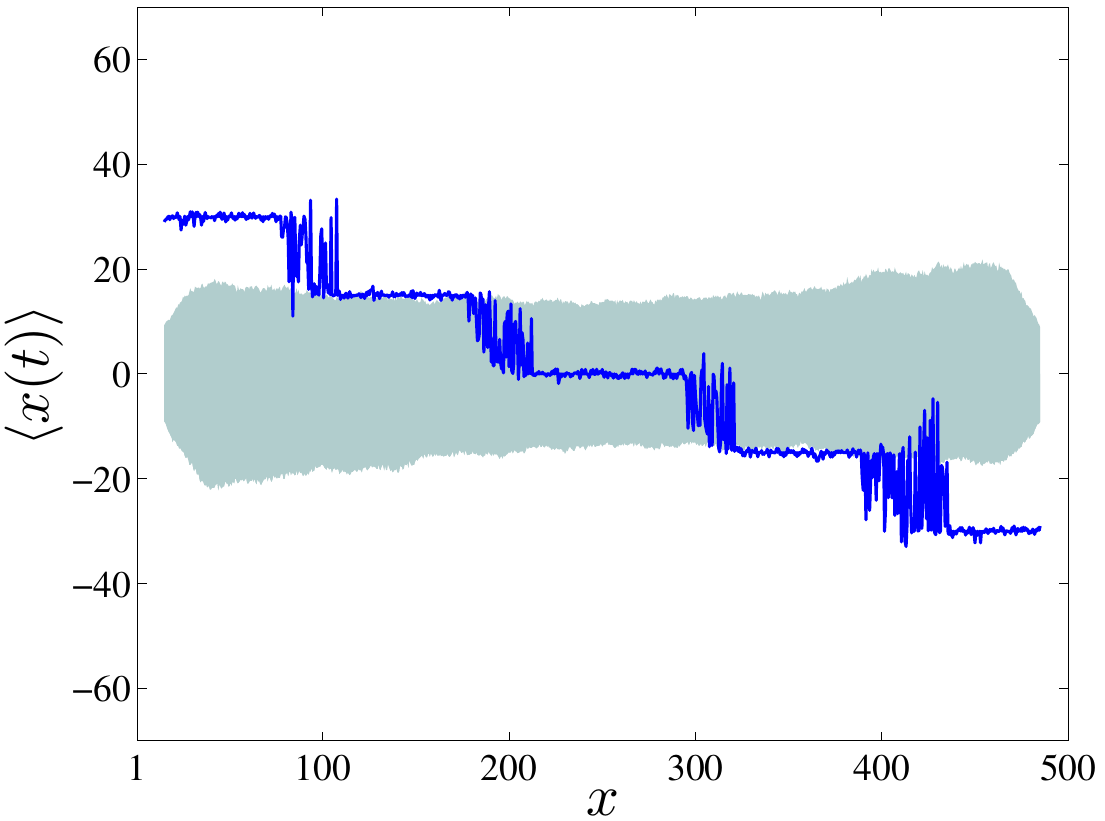}
  \raisebox{3cm}{\textsf{B}}
  \includegraphics[width=4.5cm]{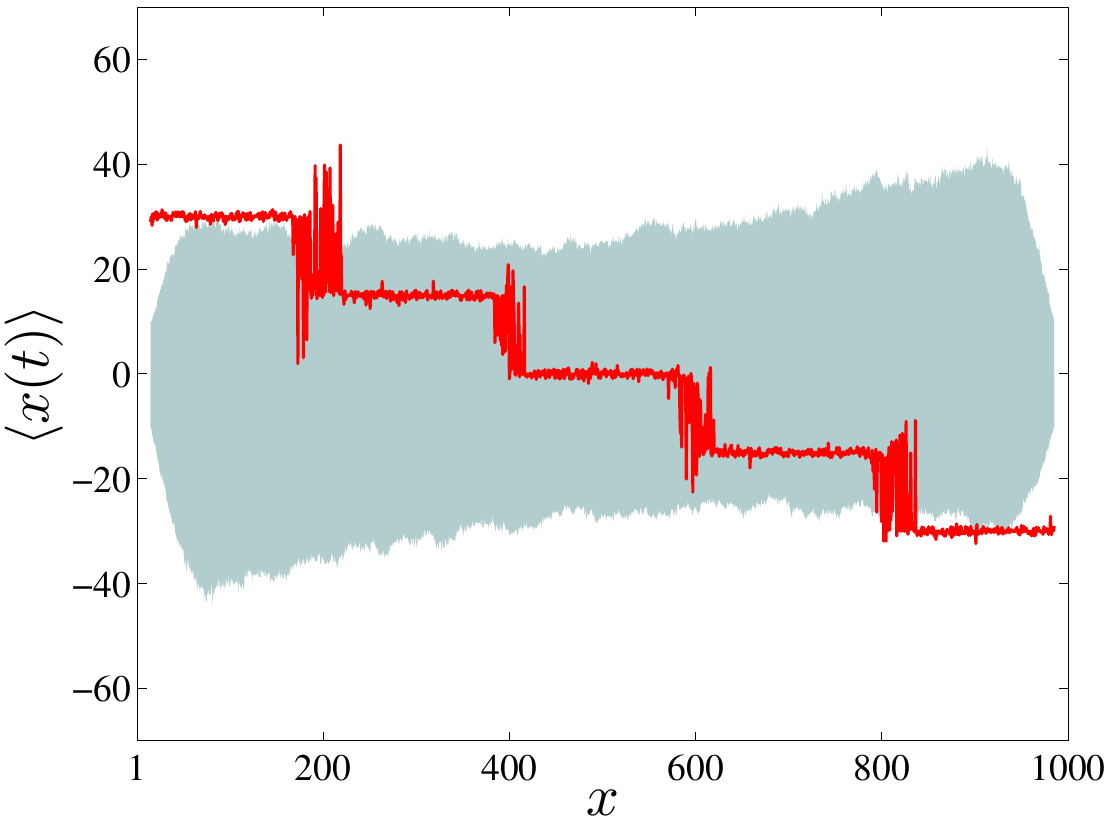}
  \raisebox{3cm}{\textsf{C}}
  \includegraphics[width=4.5cm]{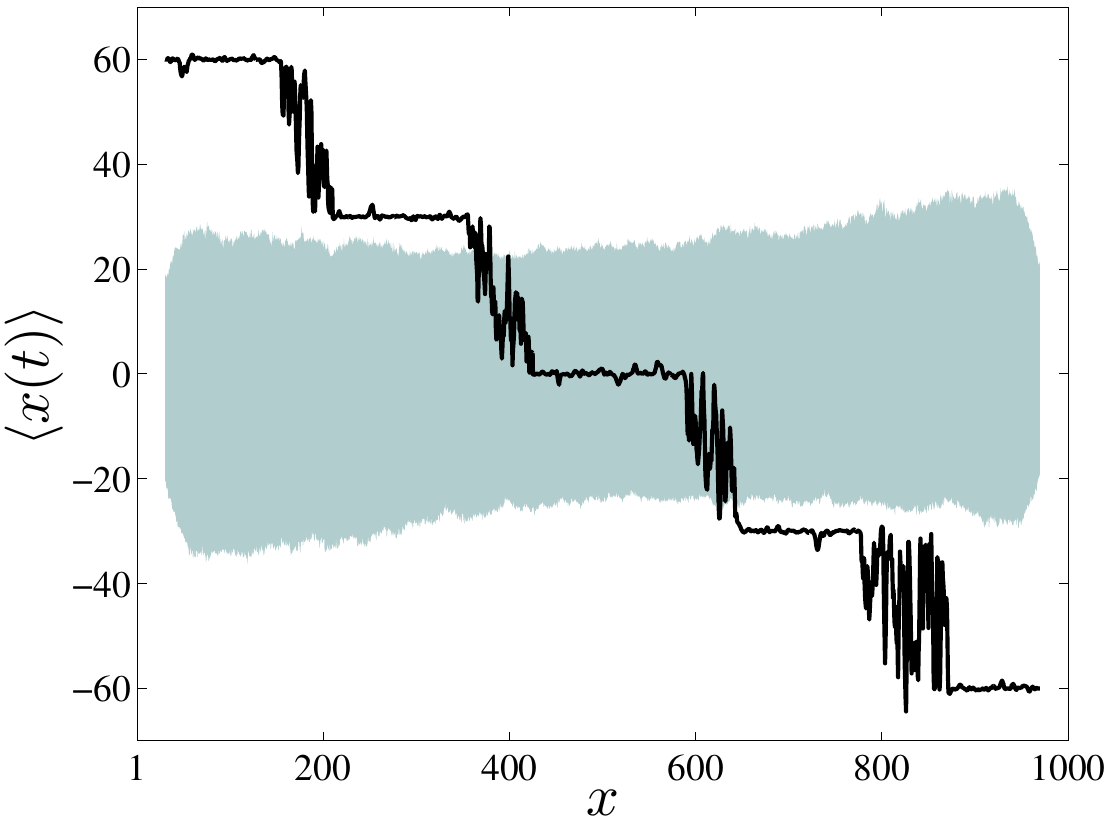}
  \caption{(color online) Thermal optimal path and the corresponding $x_{5\%} \sim x_{95\%}$ band (grey area) of Eqs.~(\ref{subEq:Synthetic A}) $\sim$ (\ref{subEq:Synthetic C}), respectively labeled by \textbf{A}, \textbf{B} and \textbf{C}. The results exhibited are obtained by the TOPS method at temperature $T=2$ and the distance definition is $\epsilon_{-}$ in Eq.~(\ref{Eq:DistMatrix}).}
  \label{Fig:BS DoubleTick}
\end{figure}

Figure \ref{Fig:BS DoubleTick} presents the thermal optimal paths and their corresponding bootstrapping significance bands obtained with the TOPS method on these three pairs of time series. Although Eq.~(\ref{subEq:Synthetic A}) and Eq.~(\ref{subEq:Synthetic B}) have ``identical'' time lags, their $x_{5\%} \sim x_{95\%}$ bands are quite different, which would lead to incorrect distinct interpretation of the two ``identical'' lead-lag signals. Moreover, even with different lead-lag structures, the bootstrapping band associated with Eq.~(\ref{subEq:Synthetic B}) and Eq.~(\ref{subEq:Synthetic C}) are quite similar due to the fact that the corresponding time series have the same duration. In addition, the three $x_{5\%} \sim x_{95\%}$ bands reject the existence of the intermediate lead-lag signal $Y(i)=aX(i)+\eta$, which actually is present and corresponds to a strong synchronization signal, which ought not to be treated as ``random''.

\subsection{Free energy test and $p$-value $\rho$}

To address the problems described in the previous section, we develop a novel criterion based on the free energy in order to assess the statistical significance of obtained thermal optimal paths. Recall that the free energies of TOP and TOPS are respectively given by expression (\ref{Eq:FreeEnergy}) and (\ref{Eq:FreeEnergy TOPS}). The key idea is to qualify a given lead-lag path by how much its free energy is statistically smaller than those of other paths that would be found in absence of a genuine signal. By construction, the selected lead-lad path has the lowest free energy, and we want to quantify how this can be quantified in a statistical test.

Consider the following synthetic lead-lag structure:
\begin{equation}
Y(i) =
\left\{
   \begin{array}{lrl}
    aX(i-\tau_1)+\eta, ~~~~~&1 &\leq i \leq 100\\
    aX(i-\tau_2)+\eta,&100 &< i \leq 200\\
    aX(i-\tau_3)+\eta,&200 &< i \leq 300\\
    aX(i-\tau_4)+\eta,&300 &< i \leq 400\\
    aX(i-\tau_5)+\eta,&400 &< i \leq 500
   \end{array}
\right.
\label{Eq:Synthetic Lead-Lag Model MS}
\end{equation}
The time series $X(t_X)$ is again generated by the first-order auto-regressive process (\ref{Eq:FirstOrder Auto-regressive}), where $b=0.7$ and $f=\sigma_{\eta}/\sigma_{\xi}=0.2$. These values ensure a strong lead-lag correlation structure between $X(t_X)$ and $Y(t_Y)$.

\begin{figure}[!ht]
\centering
  \includegraphics[width=7cm]{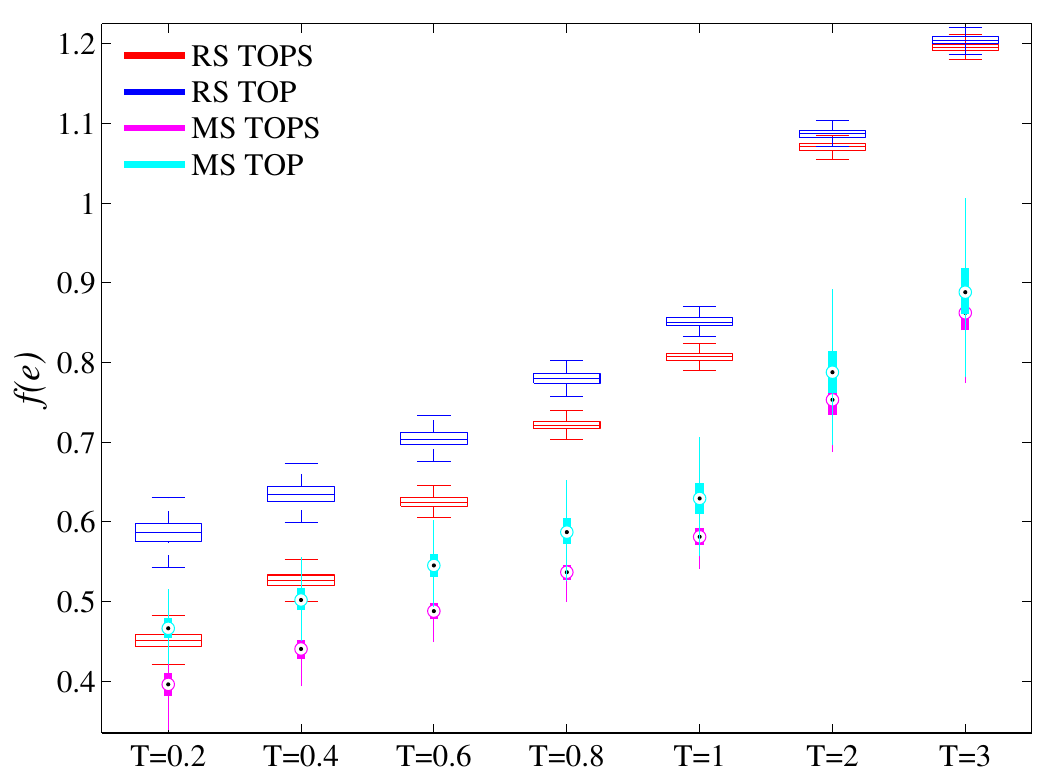}
  \caption{(color online) Boxplot of path free energies of the numerical experiments described in the text for different temperatures $T$. The color-filled boxes represent the $1000$ free energies associated with the lead-lag paths reconstructed from time series generation by Eq.~(\ref{Eq:Synthetic Lead-Lag Model MS}) (meaningful signals: MS) by using the TOP (\textcolor{cyan}{$\circ$}) / TOPS (\textcolor{magenta}{$\circ$}) method. The hollow boxes show the $1000$ free energies for the random reshuffled time series (random series: RS) using the TOP (\textcolor{blue}{$\square$}) / TOPS (\textcolor{red}{$\square$}) method. Note that the box plots obtained with the TOPS method are systematically lower than those obtained with the TOP method, confirming the superiority of the former in determining the optimal lead-lag structures.}
  \label{Fig:FE AR MPMS}
\end{figure}

In order to explore sufficiently many different lead-lag structures, we generate 1000 pairs of ($X(t_X), Y(t_Y)$) using Eq.~(\ref{Eq:Synthetic Lead-Lag Model MS}) with the lead-lag integer times $\tau_j, (j=1,2,3,4,5)$ randomly chosen in the interval $[-30,30]$ and $a$ randomly drawn in the interval $[0.7,1]$. Analyzing each of these 1000 pairs with the TOP and TOPS method, we record their free energy $e_T$ for different temperatures $T$. Then, we shuffle these time series many times and do the same analysis, recording the corresponding free energies under identical temperatures $T$. This allows us to construct 1000 distributions of random free energies associated with the 1000 meaningful series (MS), each distribution being constructing from random time series that are as close as possible from the meaningful series except for the reshuffling. As shown by the boxplots of the free energies in Fig.~\ref{Fig:FE AR MPMS}, the free energies of the meaningful time series (MS) are distinctly smaller than those for the random reshuffled time series. Moreover, we observe a systematic downward translation of the path free energies obtained by the TOPS method compared with the TOP method, confirming the superior value of the former.

\begin{figure}[!ht]
\centering
  \raisebox{3cm}{\textsf{A}}
  \includegraphics[width=4.5cm]{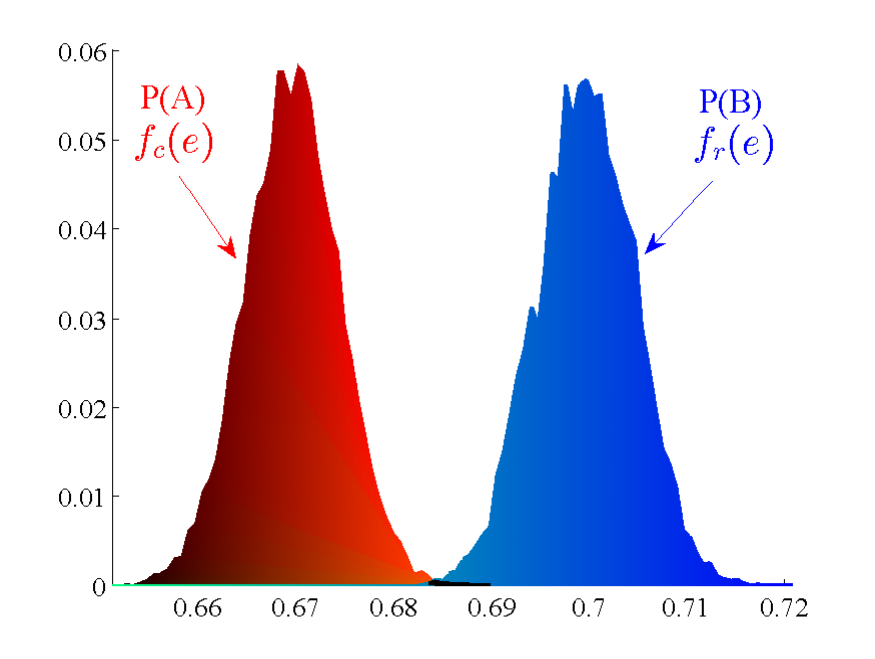}
  \raisebox{3cm}{\textsf{B}}
  \includegraphics[width=4.5cm]{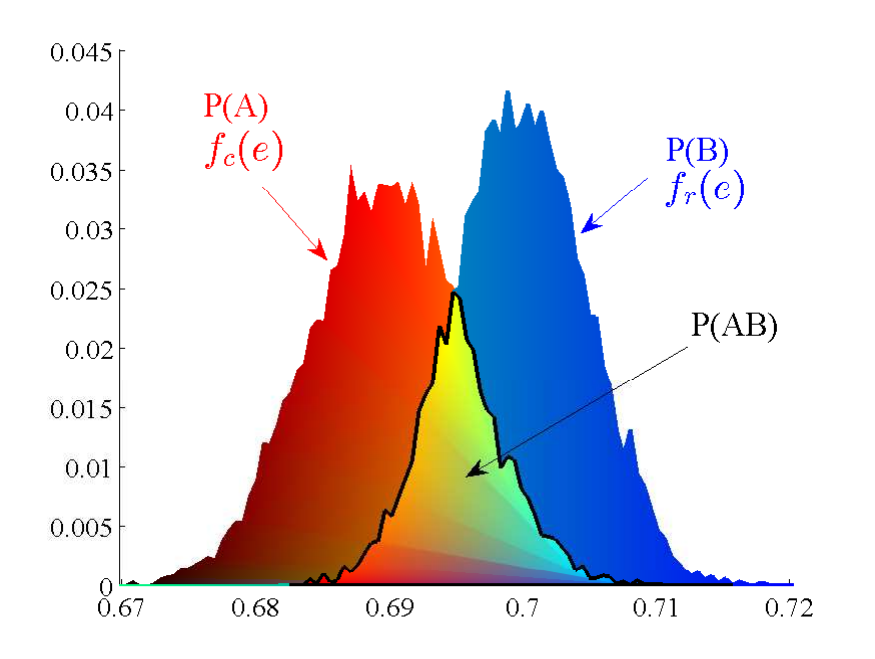}
  \raisebox{3cm}{\textsf{C}}
  \includegraphics[width=4.5cm]{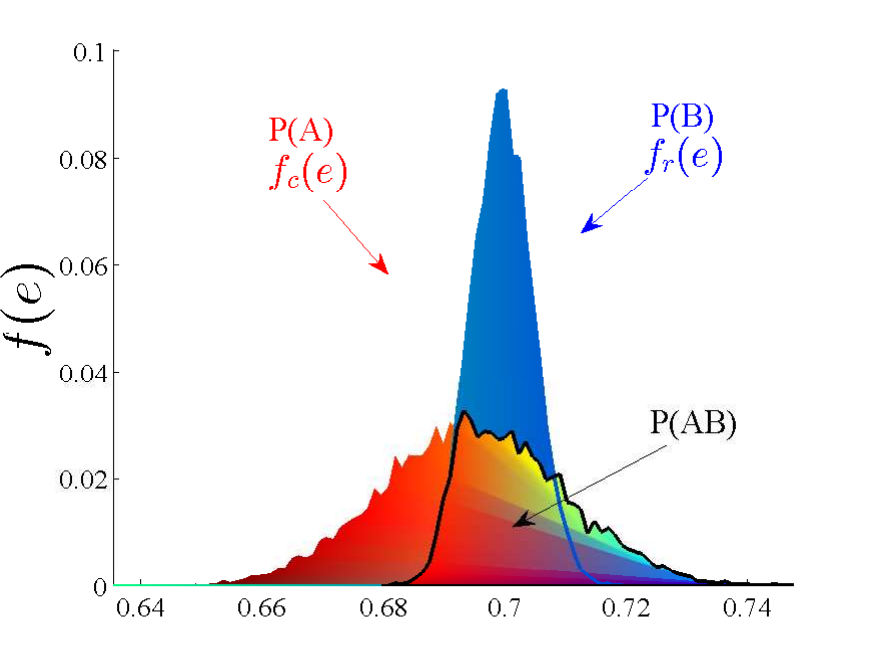}
  \caption{(color online) Schematic illustration of the construction of the free energy metric $\rho$, which can be interpreted as a p-value of the statistical test in terms of the path free energy. $P(A)$ (resp. $P(B)$) represents the total probability of the distribution $f_{c}(e)$ (resp. $f_{r}(e)$) of free energies of the lead-lag optimal paths over the 1000 meaningful signals (resp. for the random reshuffled time series), which is equal to $1$ when $f_{c}(e)$ (resp. $f_{r}(e)$) is properly normalized. Panel A (resp. C) shows a case where the recovered lead-lag structure is highly significant (resp. insignificant). Panel B depicts an intermediate situation for which it is more difficult to conclude with certainty about the reality of the found lead-lag pattern.}
  \label{Fig:FE Test Schametic}
\end{figure}

To develop the free energy selection criterion, we construct the normalized distribution $f_{c}(e)$ of free energies of the lead-lag optimal paths over the 1000 meaningful signals described above (red distribution $f_{c}(e)$ in Fig.~\ref{Fig:FE Test Schametic}). The normalized  distribution $f_{r}(e)$ of free energies of the optimal paths for the random reshuffled time series is shown with the blue color in Fig.~\ref{Fig:FE Test Schametic}. The more $f_{c}(e)$ is translated to the left of $f_{c}(e)$ and the least it overlaps with it, the more significant is the recovered lead-lag structure. This can be quantified by a simple metric $\rho$ defined as the area $P(\mathbb{B}|\mathbb{A})$ of the intersection of the two normalized  distributions, delineated with black outline in Fig.~\ref{Fig:FE Test Schametic} (\textbf{B}):
\begin{equation}
  \rho=P(\mathbb{B}|\mathbb{A}),
  \label{Eq:Overlapping Ratio}
\end{equation}
The metric $\rho$ has the simple interpretation of being the probability that a given lead-lag structure with its minimum free energy could be obtained from random time series with similar characteristics except of the lead-lag information. Thus, $\rho$ is the p-value of the statistical test in terms of the path free energy. A small value of $\rho$ indicates a strong lead-lag signal, which can be distinguished from random realization with large significance. In contrast, when $\rho$ is too large, typically $\rho > 0.2$, the lead-lag structure is obscured and not distinguishable from a random pattern at the 80\% confidence level.

\begin{figure}[!ht]
\centering
  \includegraphics[width=7cm]{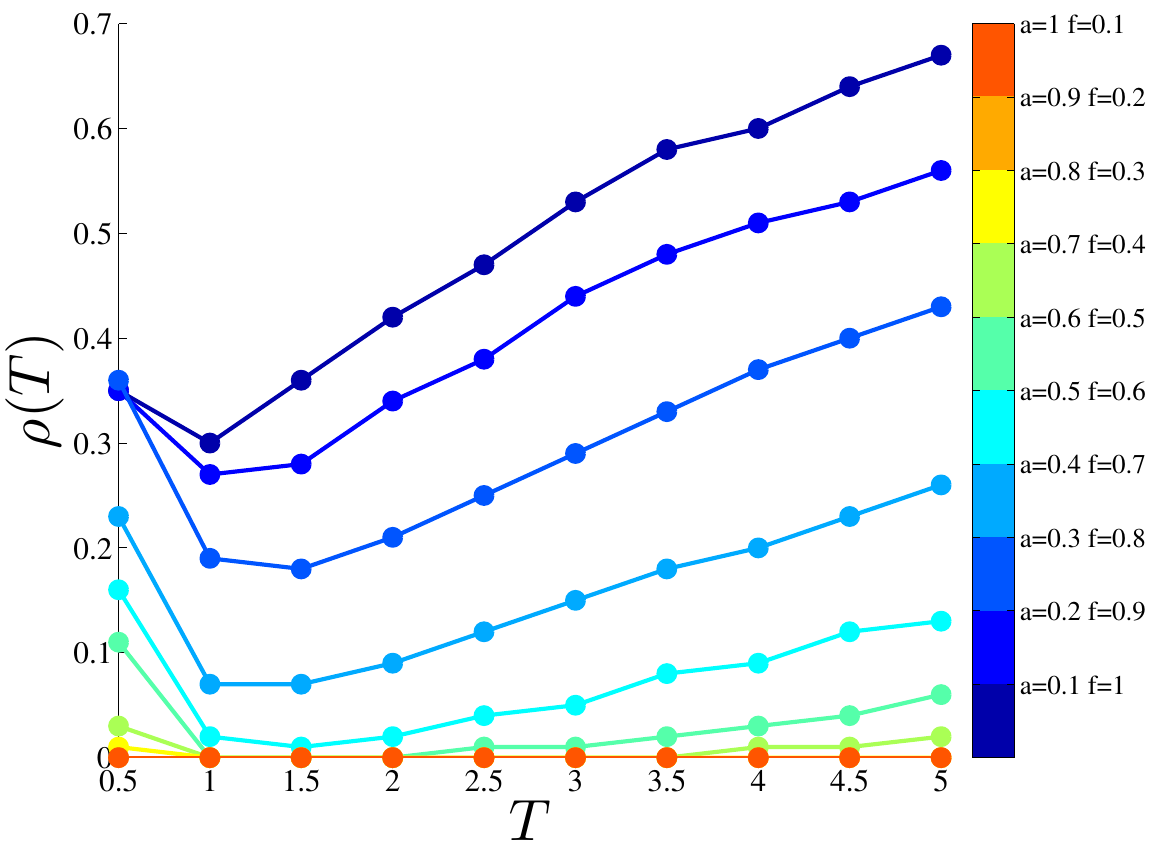}
  \caption{(color online) Free energy metric $\rho$ defined by Eq.~(\ref{Eq:Overlapping Ratio}) as a function of temperature for the ten pairs $(a,f)$ given on the right color scale and for time series generated with Eq.~(\ref{Eq:Synthetic Lead-Lag Model MS}) as explained in the text.}
  \label{Fig:FE AR Signal Test}
\end{figure}

To explore the performance of the metric $\rho$, we construct a series of statistical time series using Eq.~(\ref{Eq:Synthetic Lead-Lag Model MS}) for ten different pairs of values of the parameters $a$ and $f=\sigma_{\eta}/\sigma_{\xi}$ that control the strength of the lead-lag signals between $X(t_X)$ and $Y(t_Y)$. These ten pairs are $(a=0.1,f=1)$ (weakest interdependence and lead-lag signal), $(a=0.2,f=0.9)$, $(a=0.3,f=0.8)$, $(a=0.4,f=0.7)$, $(a=0.5,f=0.6)$, $(a=0.6,f=0.5)$, $(a=0.7,f=0.4)$, $(a=0.8,f=0.3)$, $(a=0.9,f=0.2)$ and $(a=1,f=0.1)$ (strongest lead-lag signal). For each pair $(a,f)$, we generate $1000$ pairs $(X(t_X), Y(t_X))$ by taking $b=0.7$ and $\tau_j, (j=1,2,3,4,5)$ randomly drawn in the interval $[-30,30]$. We also generate $1000$ pairs of random time series by reshuffling the time series $X(t_X)$ and $Y(t_X)$. We then calculate the free energies $e_T$ of the optimal lead-lag paths for these ten pairs $(a,f)$ using the TOPS method and obtain their corresponding free energy metric $\rho$, which are reported in Fig.~\ref{Fig:FE AR Signal Test}. For strong signals (large $a$'s and relatively small $f$'s), $\rho$ is very small and the lead-lag structure is highly significant at all explored temperature. As the signal strength weakens ($a$ decreases and $f$ increases), the choice of the temperature becomes more important and the significance also decreases. A good choice of temperature seems here to be in the range $[1,2]$, which we will make more precise below.

\begin{figure}[!ht]
\centering
  \includegraphics[width=16cm]{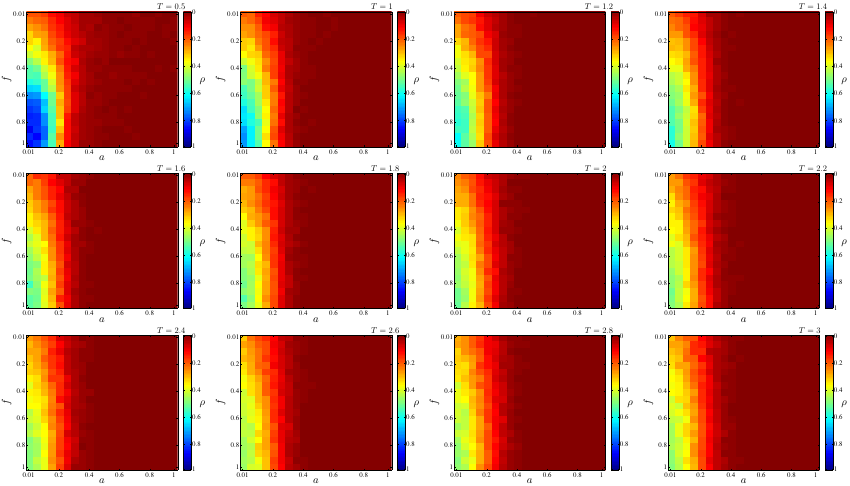}
  \caption{(color online) Maps of the signal strength $\rho$ for various $(a,f)$ combination lead-lag structures. $\rho$ close to one (blue) indicates that the corresponding lead-lag structures have weak signals and low confidence. $\rho$ close to zero (red) indicates that the lead-lag structures have strong signals and reliability. Maps are generated under different temperatures $T = 0.5, 1, 1.2, 1.4, 1.6, 1.8, 2, 2.2, 2.4, 2.6, 2.8, 3$. The $a - f$ grids are expanded by $a$ and $f$ from $0.01$ to $1$ with step $0.05$.}
  \label{Fig:Signal Strength Map}
\end{figure}

Figure \ref{Fig:Signal Strength Map} presents a more extensive view of the dependence of the free energy metric $\rho$ for 12 different temperature from $0.5$ to $3$ and for 100 different values of the parameter pairs $(a,f)$ scanning the range from $0.1$ to $1$ for both parameters in unit step of $0.05$. One can observe that the significance of the lead-lag signals is dominated by the parameter $a$. For $a > 0.5$, $\rho$ remains consistently below $0.1$ even for large additive noise quantified by $f$. However, for $a < 0.5$, the relative noise amplitude $f$ begins to impact the $\rho$ values.

\begin{figure}[!ht]
\centering
  \includegraphics[width=7cm]{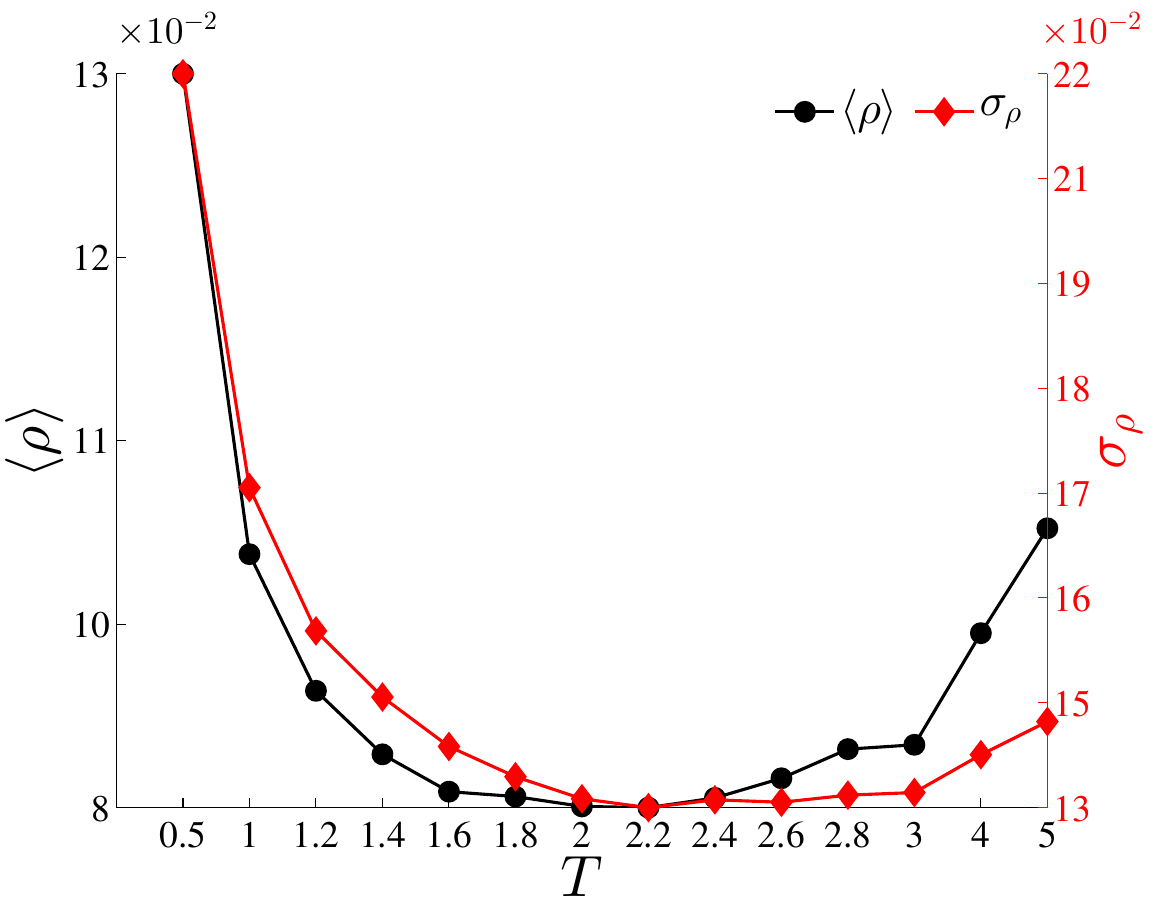}
  \caption{(color online) Mean values and standard deviations of $\rho$ values obtained in the signal maps of figure \ref{Fig:Signal Strength Map} as a function of temperature $T$.}
  \label{Fig:Signal Map Means Deviations}
\end{figure}

As already identified in Fig.~\ref{Fig:FE AR Signal Test}, the choice of the temperature $T$ used in the implementation of TOPS plays a useful role in the tradeoff between ``not extracting too much from the spurious noise'' and ``washing out too much of the relevant signal''. Too small values of $T$ make the TOPS paths very sensitive to $a$ and $f$, resulting in large fluctuations in $\rho$, which becomes unreliable. Too large values of $T$ can obscure the relevant signal since the entropy of the path configurations dominates over the minimization of the distance. This leads to large insignificant values of $\rho$. Fig.~\ref{Fig:Signal Map Means Deviations} shows the mean values and standard deviations of $\rho$ obtained in the signal maps of Fig.~\ref{Fig:Signal Strength Map} as a function of temperature $T$. One can observe a minimum of the mean value and of the standard deviation almost at the same temperature close to $T = 2.2$. This supports our choice below to use a temperature close to $T=2$ in order to optimize the recovery of the optimal lead-lag paths.

\subsection{Impact of temporal dependence}

It is common that many economic and financial time series contain serial autocorrelations. For instance, we find that the HPI return time series have significant autocorrelations, especially for the US HPI. The reshuffling procedure adopted in the previous subsection destroys not only the lead-lag relations between the two time series, but also the temporal dependence within each of the individual time series. An important issue is about the impact on the free energy test of serial autocorrelations in the two time series under investigation. If there is substantial memory (e.g. evidenced by autocorrelation across several lags) this should also be clearly reflected by the bootstrap data in order to obtain a reasonable reference distribution under the null hypothesis of no lead-lag relations between the two series. This could be achieved by using the stationary bootstrap proposed by \cite{Politis-Romano-1994-JASA}, which takes into account serial dependence. In their original stationary bootstrap approach, the data are resampled with replacement. However, in our free energy test, the original data should be kept totally unchanged except for their sequence. Thus, we make a revision to the stationary bootstrap so that the data can be resampled without replacement.

We repeat the test for the mathematical model used in Eq.~(\ref{Eq:Synthetic Lead-Lag Model MS}). The procedure is described as follows. The subscript sequence $\{1,2,\cdots,N\}$ is divided into a sequence of $n$ consecutive non overlapping blocks $B_k=\{\sum_{j=1}^kL_{j-1}+1, \cdots, \sum_{j=1}^kL_j\}$ with $k=1,\cdots,n$, $L_0\equiv0$ and $\sum_{j=1}^nL_j=N$, where $L_k$ (except for $L_0$) is a sequence of independent and identically distributed random variables drawn from the geometric distribution with fixed parameter $p\in[0,1]$. In practice, one can generate a sequence of random numbers $\{l_1, l_2, \cdots, l_{n-1}, l_n, \cdots\}$ and take $\{l_1, l_2, \cdots, l_{n-1}, N-\sum_{j=1}^{n-1}l_j\}$ as a realization of $\{L_1, L_2, \cdots, L_n\}$, in which $n$ is determined as the minimum number satisfying $N-\sum_{j=1}^{n-1}l_j\leq l_n$. The two time series $X(i)$ and $Y(i)$ are thus divided into $n$ non overlapping blocks. We block shuffle $X(i)$ and $Y(i)$ {\textit{simultaneously}} to obtain $X^*(j)$ and $Y^*(j)$ by shuffling the subscript blocks $B_k$ so that the original subscripts of the resulting time series are the same for $X$ and $Y$.

\begin{figure}[!ht]
\centering
  \includegraphics[width=7cm]{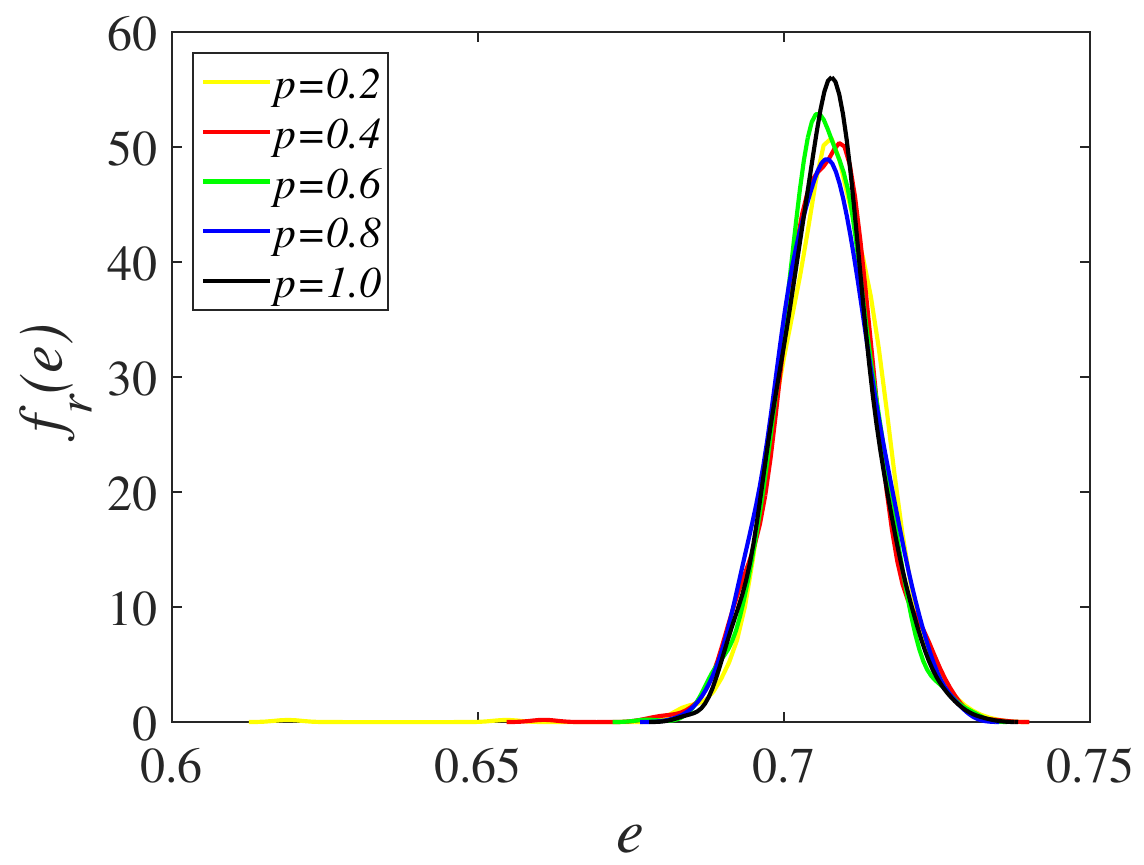}
  \caption{(color online) Impact of serial autocorrelations in the time series under investigation on the free energy test.}
  \label{Fig:FreeEnergy:BlockBootstrap}
\end{figure}

We generate $1000$ pairs of block reshuffled random time series $X^*$ and $Y^*$ for each $p=[0.2,0.4,0.6,0.8,1.0]$ and then calculate the free energies $e_T$ of the TOPS paths. We plot the distributions of $e_T$ with different $p$ in Fig.~\ref{Fig:FreeEnergy:BlockBootstrap}. We show that the block shuffled cases with serial autocorrelations ($p<1$) have almost the same free energy distributions as the random shuffled case ($p=1$). We could give some intuitive explanation for the little influence of temporal dependence from the tradeoff perspective. When we conduct TOPS analysis for the block reshuffled series, for each block, the weighted sum of local energy maybe smaller in probability than that for the same length subseries by random reshuffling, due to the retained consistency in the block. However, from another aspect, the block reshuffling approach can cause big cliff between two nearby blocks. This will produce higher energy because the lead-lag path is required to be continuous and cannot be fully adjusted immediately. Therefore, the global energy for one path may have no obvious difference with that of random reshuffling due to the tradeoff between these two opposite effects. And then, the free energy distribution naturally shows nothing significantly different. Therefore, we conclude that serial autocorrelations in the raw time series do not have evident impacts on the the free energy test. Hence, in the rest of this work, we will adopt the random shuffling approach to reduce computational time.

\subsection{Self-consistent test of the lead-lag structure}

In addition to the p-value introduced in the previous section, we present the self-consistent test for lead-lag structures identified by the TOPS method and qualified by the metric $\rho$. Its idea is straightforward: for the lead-lag path $\langle x(t) \rangle$ to be significant, synchronizing the two time series using the time varying $\langle x(t) \rangle$ should lead to a statistically significant correlation.

Consider two standardized time series $X(t)$ and $Y(t)$ for which the TOPS method has determined the optimal thermal lead-lag path $\langle x(t) \rangle$. If $\langle x(t) \rangle$ is indeed to be interpreted at the lead-lag time between the two time series, this implies that $X(t-\langle x(t) \rangle)$ and $Y(t)$ should be synchronized and should exhibit a strong linear dependence. In other words, forming the regression
\begin{equation}
  Y(t)=c+a X(t-\langle x(t) \rangle)+\varepsilon(t)~,
  \label{Eq:Synchronized Self-Consistent}
\end{equation}
the coefficient $a$ should be significantly different from $0$.

Let us illustrate this by using the two time series $X(t_X)$ and $Y(t_Y)$ generated by Eqs. (\ref{Eq:Synthetic Lead-Lag Model}) and (\ref{Eq:FirstOrder Auto-regressive}). The corresponding thermal optimal path $\langle x(t) \rangle$ quantifying the lead-lag relationship between $X(t_X)$ and $Y(t_Y)$ has been explored and exhibited in Fig.~\ref{Fig:TOPS TOP Path AR}. We apply the transformation $X(t) \to X(t-\langle x(t) \rangle)$ and perform the regression (\ref{Eq:Synchronized Self-Consistent}) to obtain the estimate $\hat{a}$ of the regression coefficient $a$. Applying a standard t-test at the $95\%$ confidence level allows one to conclude if $\hat{a}$ is statistically different from $0$ and decide if there is a statistically significant correlation between the first translated time series $X(t-\langle x(t) \rangle)$ and the second time series $Y(t)$.

In order to evaluate the free energy metric $\rho$, we estimate the amplitude of the residuals $\varepsilon(t)$ via the metric
\begin{equation}
  \hat{f}=\frac{\sqrt{\sum^N_{k=1}[Y(k)-\hat{a}X(k-\tau(k))]^2}}{\sqrt{\sum^N_{k=1}(X(k)-\bar{X})^2}}
\end{equation}
where $\bar{X}$ is the mean value of $X(t_X)$ and $N$ is the length of time series ($N=500$ in this case). Using the estimate $\hat{a}$ of the regression coefficient in expression (\ref{Eq:Synchronized Self-Consistent}) and its corresponding average residual amplitude $\hat{f}$, we can read off Fig.~\ref{Fig:Signal Strength Map} for $T=2$ the value of $\rho$ corresponding to the discrete set closest to the found pair $(\hat{a}, \hat{f})$. Alternatively, following the procedure used to construct Fig.~\ref{Fig:Signal Strength Map}, we can generate many synthetic lead-lag structures with $(a, f)=(\hat{a}, \hat{f})$ and the recovered lead-lag path $\langle x(t) \rangle$ and compare their free energies with the corresponding random cases.

If the estimated regression coefficient $\hat{a}$ is found statistically significant from zero and if the free energy p-value $\rho$ is small, we should conclude that the found lead-lag path is genuine.

\begin{figure}[!ht]
\centering
  \includegraphics[width=8cm]{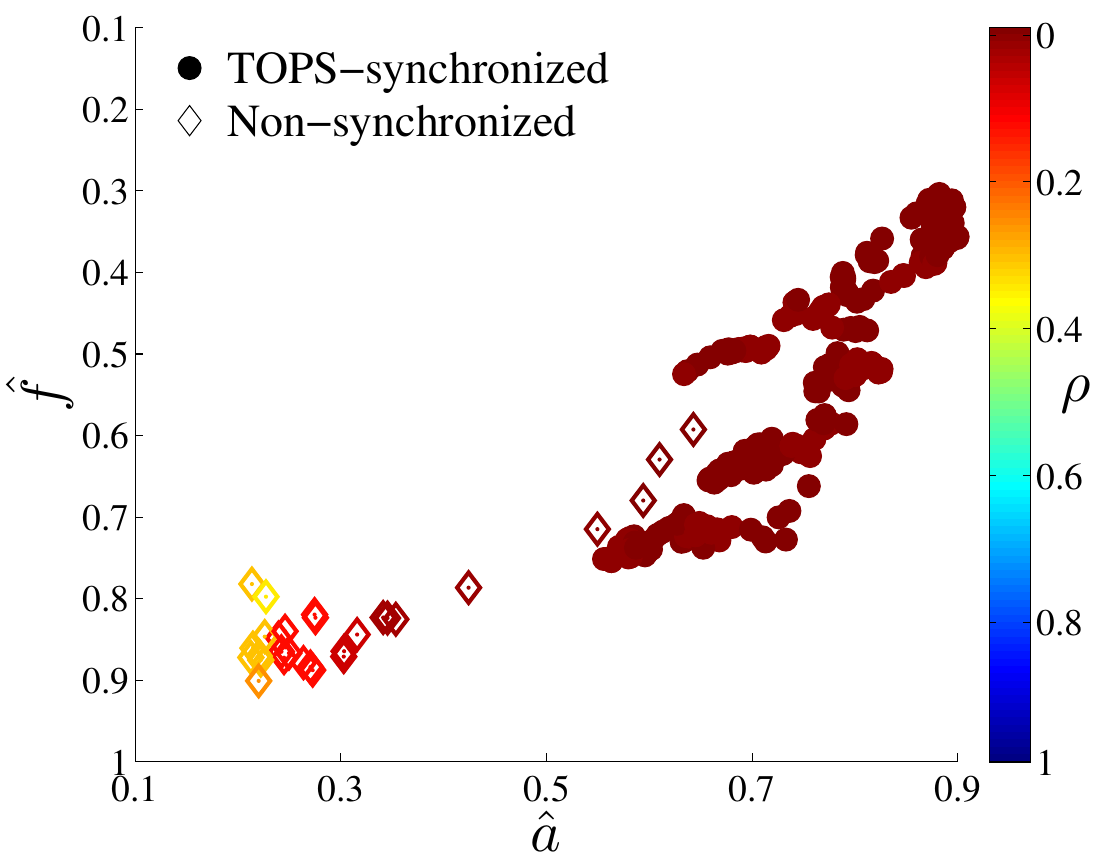}
  \caption{(color online) Combining the free energy p-value $\rho$ criterion and the self-consistent test of the statistical significant of the regression coefficient $a$ of the second time series $Y(t)$ regressed with respect to the shifted first time series $X(t-\langle x(t) \rangle)$ using the obtained lead-lag structure derived with the TOPS method with temperature $T=2$. Each symbol shows the values of the pair ($\hat{a},\hat{f}$) obtained in a moving window of size $n=100$ and the color encodes the $\rho$ values in the $T=2$ signal strength map of Fig.~\ref{Fig:Signal Strength Map}. The filled circles show the estimations of the TOPS-synchronized series, namely $X(t-\langle x(t) \rangle)$ and $Y(t)$. The empty diamonds are the estimations of the non-synchronized series, namely $X(t)$ and $Y(t)$.  The original synthetic time series have been generated with  $a=0.7$ and $f=0.2$.}   \label{Fig:AR Singla Moving Window}
\end{figure}

Figure \ref{Fig:AR Singla Moving Window} shows the result of the joint application of the self-consistent test and of the free energy p-value $\rho$ criterion to the time series generated by  Eqs. (\ref{Eq:Synthetic Lead-Lag Model}) and (\ref{Eq:FirstOrder Auto-regressive}). We implemented the tests in moving windows with size $n=100$ and moving forward with step of one time unit, leading to a total of $N-n+1=500-100+1=401$ windows. Within each window, we synchronized (resp. did not synchronized) the time series, estimating the coefficients $\hat{a}$ and $\hat{f}$, and read off the corresponding $\rho$ values  on the signal strength map shown in Fig.~\ref{Fig:Signal Strength Map}, for $T=2$. Among the $401$ TOPS-synchronized windows, there are $358$ windows that have statistical significant coefficient $\hat{a}$. In contrast, for the $401$ overlapping non-synchronized windows, there are only $33$ windows that pass the $t$ test of significant non-zero coefficient $a$. The lead-lag signals of the TOPS-synchronized windows are also remarkably stronger than for the non-synchronized windows. Most of the TOPS-synchronized windows are found in the region of the signal strength map with very small $\rho$ values, while most of the significant non-synchronized windows are located within $0.15 < \rho < 0.4$ and cannot be distinguished from random lead-lag patterns. The original synthetic time series have been generated with  $a=0.7$ and $f=0.2$ and we can observe a cluster of estimated  $(\hat{a},\hat{f})$ of the TOPS-synchronized windows in the neighborhood of these values. One can note that the method tends to overestimate the noise amplitude.

\section{Application to house prices and monetary policy}
\label{Sec:Application}

\subsection{Data description}

In the following, we apply the symmetric TOPS method, together with the free energy p-value $\rho$ criterion and self-consistency test of the lead-lag structure introduced in the previous section to analyze the time dependent lead-lag structures of house price and monetary policy of the United Kingdom (UK) and the United States (US). For the UK, our study compares two major house price indices (HPIs) with the interest rates covering short to long term maturities including the Bank of England's Official Bank Rate (BR), the 1-year British Pound LIBOR (BPLibor) and the 10-year Government Bond Generic Bid Yields (GBY) constructed by Bloomberg. The house price datasets used are gained from the largest mortgage providers in the UK, namely the seasonal adjusted Nationwide Building Society HPI (NBSHPI) and the seasonal adjusted Halifax HPI (HHPI). For the US, the seasonal adjusted house price indices published by the Federal Housing Finance Agency (FHFAHPI) and the Standard $\&$ Poor's Case-Shiller Index (SPCSHPI) are compared with the effective Federal Funds Rate (FFR), which is the open market rate that follows the Target Rate issued by the Federal Reserve, the 1-year dollar Libor (DLibor) and the 10-year Treasury Bill (TB). All the time series span from January 1991 to April 2011 monthly except SPCSHPI that spans from January 2000 to April 2011 monthly.

For the subsequent analysis, logarithmic returns are used for the analysis of the monthly data sets. The returns are defined as continuously compounded returns as follows:
\begin{equation}
r(t)=\ln(S(t))-\ln(S(t-1))
\label{Eq:LogarithmicReturn}
\end{equation}
where $S(t)$ is the initial non-stationary time series (HPI or monetary policy series) at time $t$. After generating the return time series $r(t)$, we then normalize all $r(t)$ according to
\begin{equation}
R(t)=\frac{r(t)}{\sqrt{\langle [r(t)]^2 \rangle}}.
\label{Eq:ReturnNormalization}
\end{equation}
Usually, we standardize the return series so that the mean of standardized returns is zero and their standard deviation is equal to one. This procedure ensures the comparability of the time series and allows to produce meaningful results during the TOP method analysis. This looked {\textit{a priori}} as quite reasonable but may be actually misleading in the current case.
Indeed, consider the following artificial example. Suppose that the FFR increases by $0.1\% + \epsilon_1$ per month and the HPI increases also by $0.2\% + \epsilon_2$ per month with a lead of 3 month, where $\epsilon_{1,2}$ are idiosyncratic noises. This corresponds to the equation
\begin{equation}
  r_{\rm{FFF}}(t) = 0.5*r_{\rm{HPI}}(t-3) + \epsilon(t),
  \label{Eq:Eg1}
\end{equation}
which exhibits a clear lag correspondence with positive correlation. Now, by standardizing the two time series, we remove the two trends (constant growth rates) and are left with two processes
\begin{equation}
  R_{\rm{FFR}}(t) = \epsilon_1(t) ~~{\mathrm{and}}~~~  R_{\rm{HPI}}(t) = \epsilon_2(t)
  \label{Eq:Eg2}
\end{equation}
with no genuine lead-lag correspondence between the two time series but just noise. This simple example illustrates the danger of the standardization procedure for this application. However, in order to make the two time series comparable in the definition of the norms, we need to normalize the returns by their standard deviations, but NOT to remove the means. Obviously, when the mean is close to zero such as for stock returns, there is no pronounced difference between standardization and normalization. By employing the Engle-Granger test and the Johansen test, we checked that there is no co-integration between interest rates and house price indices.

\begin{figure}[!ht]
\centering
  \includegraphics[width=7cm]{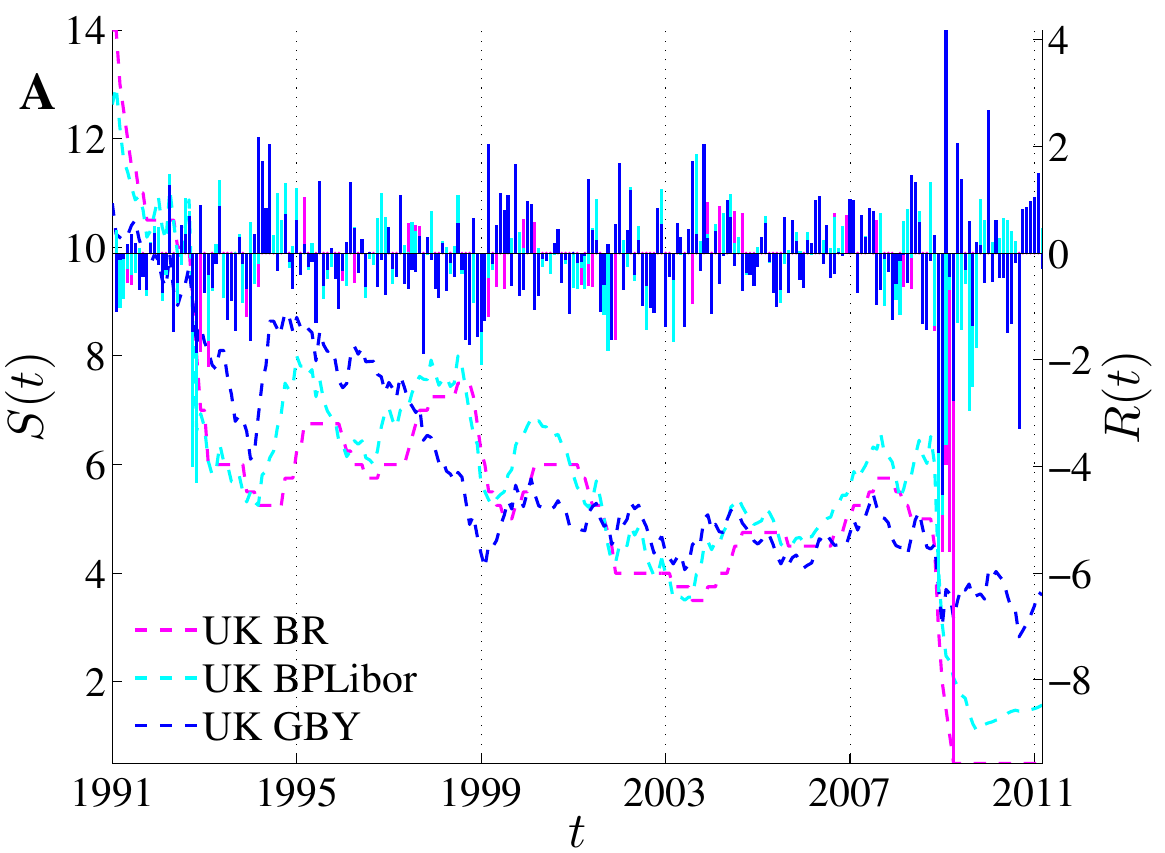}
  \includegraphics[width=7cm]{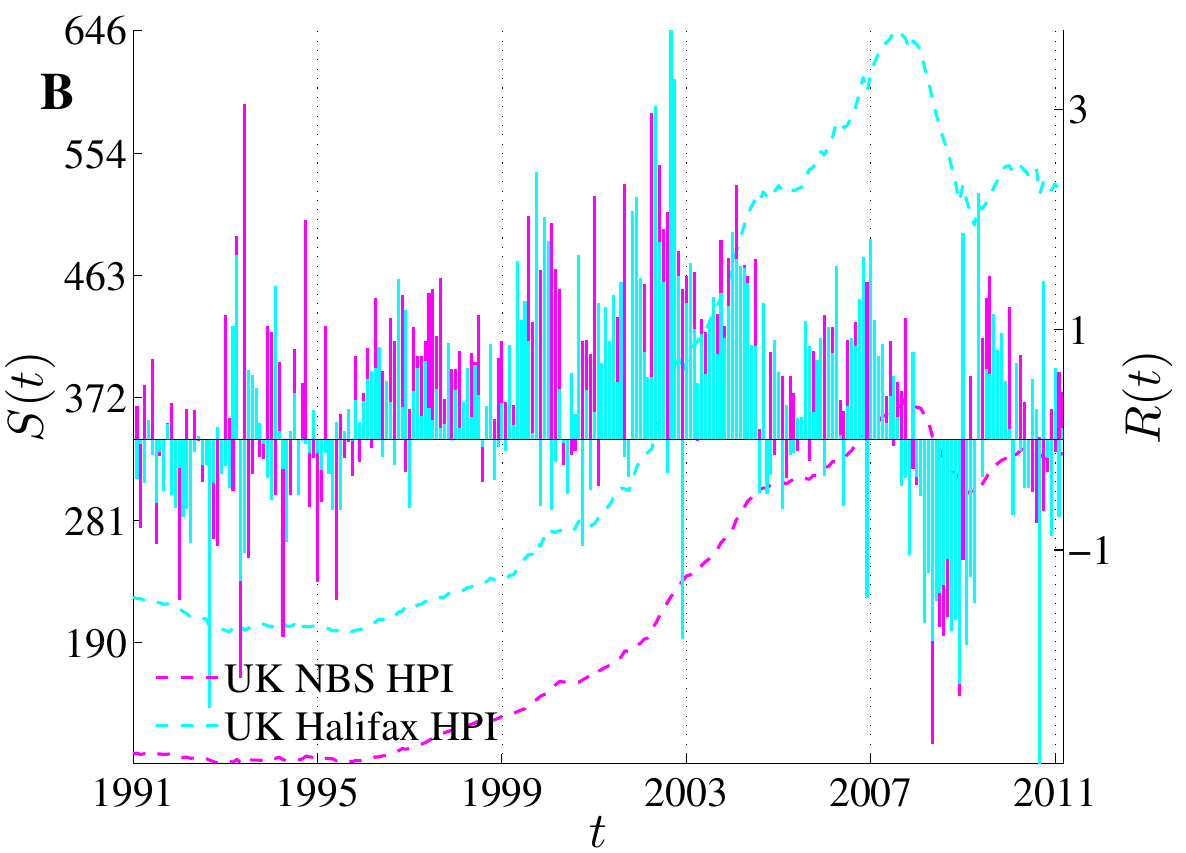}
  \includegraphics[width=7cm]{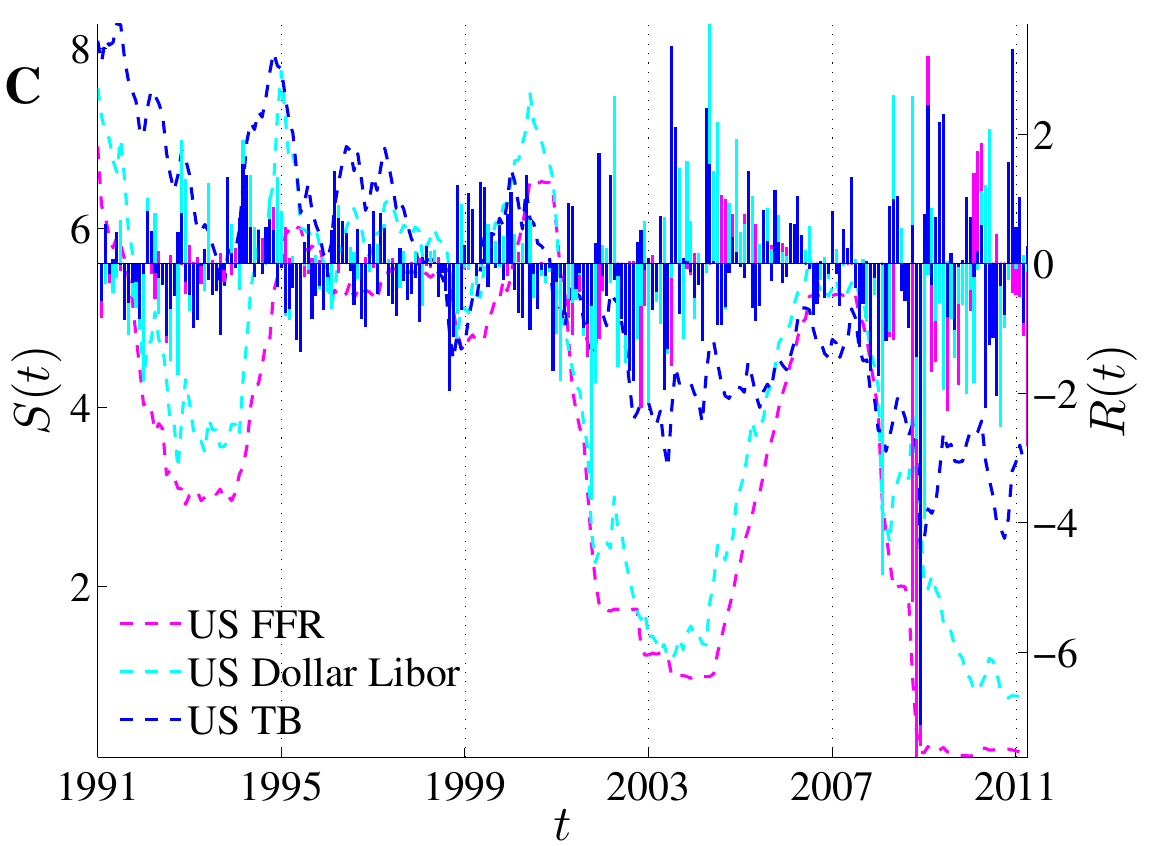}
  \includegraphics[width=7cm]{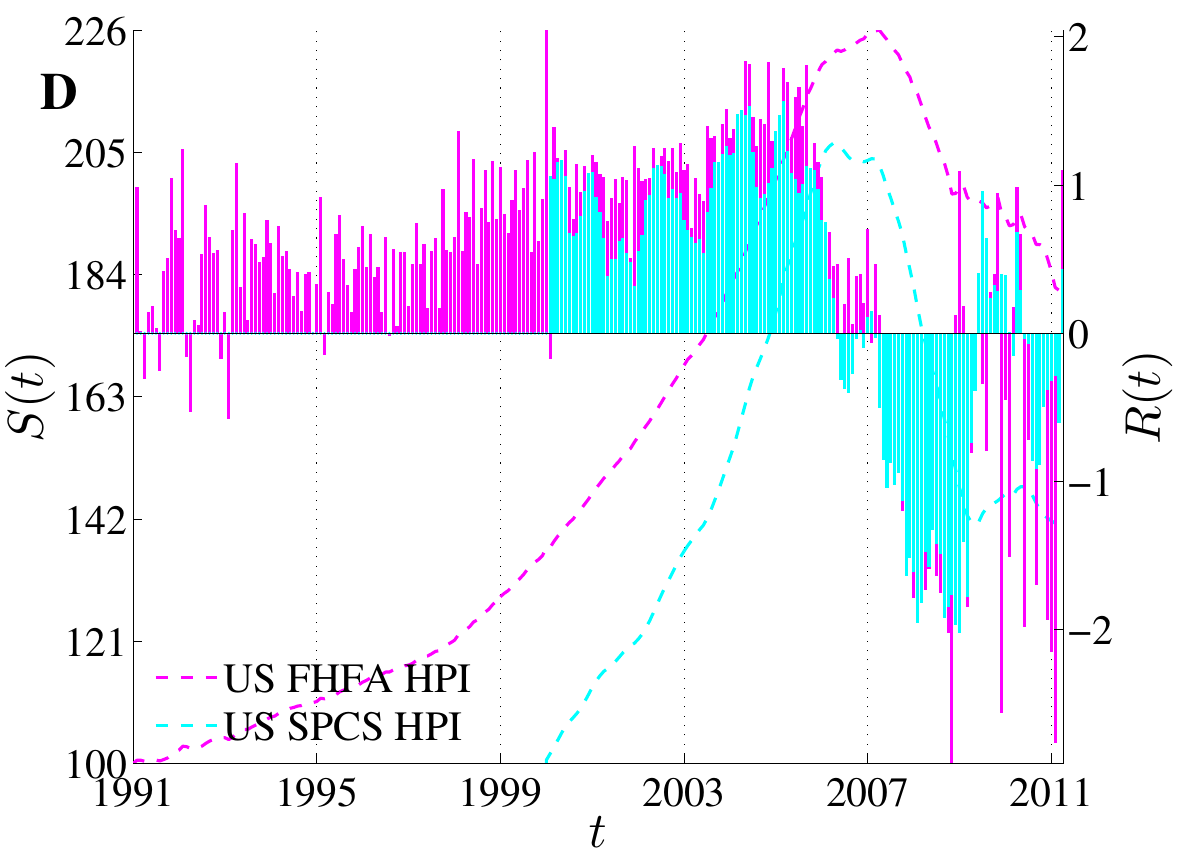}
  \caption{(color online) Data sets presentation. The left y axis $S(t)$ of (\textbf{A}) exhibits the monthly UK interest rate data of the Bank of England's Official Bank Rate (BR \textcolor{cyan}{---}), the British Pound 1-year LIBOR (BPLibor \textcolor{magenta}{---}) and the 10-year Government Bond (GBY \textcolor{blue}{---}) as well as their corresponding normalized  returns $R(t)$ at the right y axis. $S(t)$ of (\textbf{B}) is the seasonal adjusted house price index published by Nationwide Building Society (NBSHPI \textcolor{magenta}{---}) and the seasonal adjusted house price index for all house types published by Halifax (HHPI \textcolor{cyan}{---}) as well as their corresponding  normalized  return $R(t)$ at the right y axis. $S(t)$ of (\textbf{C}) exhibits the monthly US interest rate data of the Effective Federal Fund Rate (FFR \textcolor{cyan}{---}), the 1-year US Dollar LIBOR (DLibor \textcolor{magenta}{---}) and the 10-year Treasury Bond (TB \textcolor{blue}{---}) as well as their corresponding normalized  return $R(t)$ at the right y axis. $S(t)$ of (\textbf{D}) is the US house price index published by Federal Housing Finance Agency (FHFAHPI \textcolor{magenta}{---}) and the Standard $\&$ Poor's Case-Shiller house price index (SPCSHPI \textcolor{cyan}{---}) as well as their corresponding  normalized  returns $R(t)$.}
  \label{Fig:TimeSeries Initial Return}
\end{figure}

Figure \ref{Fig:TimeSeries Initial Return} (\textbf{A}) illustrates the monthly development of the UK interest rates over two decades between 1991 and 2011. Within this period, an almost continuous decline of interest rates with two particularly steep drops during 1991-1992 and by the end of 2008 is observed. In the late 1980s, the Bank of England had to cope with the after-effects of the stock market crash of October 1987 leading to comparably low interest rates, which supported the creation of a house price bubble. With the intention to stop this development, the Bank of England increased its base rate. However, in 1991 the bubble collapsed with a significant impact on the UK economy, resulting once again in a decline of interest rate levels. This development was further pronounced by the so-called ``Black Wednesday'', which was triggered by the UK joining the European Exchange Rate Mechanism (ERM). In the period between 1992 and 2008, the interest rates remained at a relatively stable level with a less pronounced downward trend and only minor fluctuations around an average interest rate of roughly $5\%$. Stronger variations only occurred around the period of the bust of the dotcom bubble in late 2000. The most recent drop in interest rate levels was directly associated with the peak in UK as well as United States house prices around 2006 and the response to the following credit crunch in 2007.

Figure \ref{Fig:TimeSeries Initial Return} (\textbf{B}) shows the seasonal adjusted development of housing prices measured by the Nationwide Building Society and the Halifax HPI. After the collapse of the 1991 housing bubble, the prices stayed almost constant for over half a decade, followed by a period of massive growth with a small plateau in the end of 2004 and a climax in late 2007. Starting from early 2009, the house prices exhibited a moderate rebound that was again followed by a small decrease of the valuation of real estate.

Figure \ref{Fig:TimeSeries Initial Return} (\textbf{C}) presents the development of US interest rates over two decades. Three periods of decreasing interest rates, led by the target rate of the Federal Reserve (FFR), can be identified. These changes in the interest rate policy were triggered by periods of economic recessions, namely the US Savings and Loans Crisis between 1990 and 1991, the dotcom bubble that climaxed and crashed in March 2000, together with the shock of the 9/11 terrorist attacks in 2001 and the most recent subprime mortgage crisis that caused a global financial crisis beginning in late 2007. Each period preceding the crisis was characterized by a sustained economic growth with slow monotonous rises of the Federal Reserve Target Rate likely aiming at tempering a possible overheating. In general, during the last two decades, the interest rate level has decreased significantly for all maturities, resulting in historical lows today.

The development of house prices in the United States, represented by the FHFA House Price Index and the S\&P's Case-Schiller House Price index in Fig.~\ref{Fig:TimeSeries Initial Return} (\textbf{D}) is characterized by a strong average exponential growth since the early 1990s (and even faster-than-exponential growth in the more recent period from 2003 to 2007 \citep{Sornette-Cauwels-2014-Risks,Leiss-Nax-Sornette-2015-JEDC}), which peaked in 2007. It is also important to note that the house price time series is obviously non-stationary and even the logarithmic returns seem to have three different regimes, one until 1997-1998 with small growth, another until 2007 with increasing growth rates and a strong decay until the end of 2010. It is interesting to see that the peak of the UK real estate prices is delayed by over one year compared to the peak in the US.

\subsection{TOPS analysis on the time dependent lead-lag structures of house price and monetary policy}

\begin{figure}[ht]
\centering
  \includegraphics[width=14cm]{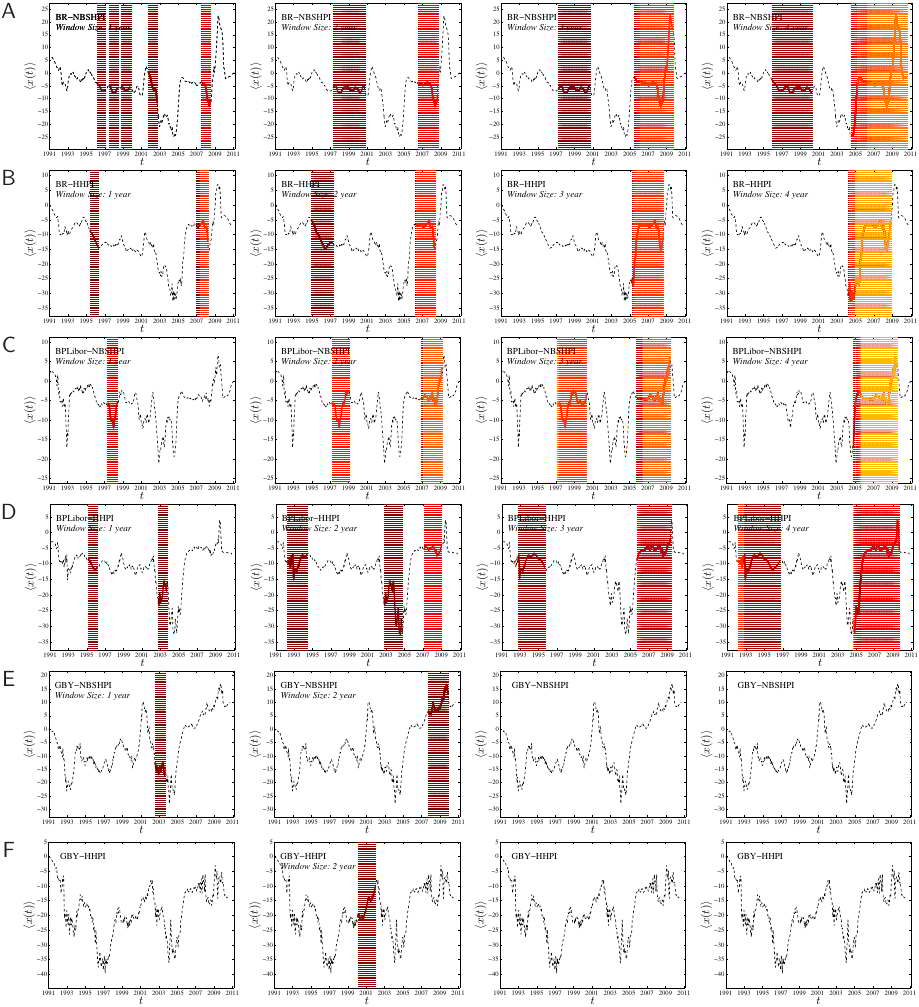}
  \caption{(color online) TOPS analysis of the  normalized  returns $R(t)$ of monthly interest rates and of seasonal adjusted house price index of the UK. The analysis is implemented at temperature $T=2$, using the distance definition $\epsilon_{-}$. Each row shows the result for a pair of ``interest rate vs. house price index'' series. The pairs  BR vs. NBSHPI, BR vs. HHPI, BPLibor vs. NBSHPI, BPLibor vs. HHPI, GBY vs. NBSHPI, GBY vs. HHPI correspond respectively to the rows \textsf{A},\textsf{B},\textsf{C},\textsf{D},\textsf{E},\textsf{F}. The self-consistent test is implemented within moving windows with sizes of $1$ to $4$ years, corresponding to columns $1$ to $4$ respectively. Each black dash line is the resulting TOPS path $\langle x(t) \rangle$, which is chosen as the one with lowest free energy among $41 \times 41$ paths of different starting points ($t_X=i_1,t_Y=i_2$) and ending points ($t_X=N-i_1,t_Y=N-i_2$) for $i_1,i_2=0,1,2,\cdots,30$. The case when $\langle x(t) \rangle > 0$ indicates that interest rate changes are  preceding house price index changes at time $t$, and vice versa. Periods qualified by the self-consistent test are highlighted by shadow areas, whose colors encode their $\rho$ values following the same convention as in the signal strength map (Fig.~\ref{Fig:Signal Strength Map}, $T=2$). The $\rho$ values are calculated based on their estimated $\hat{a}$ and $\hat{f}$ as explained in the text.}
  \label{Fig:TOPS UK}
\end{figure}

Figure \ref{Fig:TOPS UK} and Fig.~\ref{Fig:TOPS US} show the results of the TOPS analysis, together with the free energy p-value $\rho$ criterion and self-consistency test of the lead-lag structure performed on the time series presented in the previous section. For each TOPS path of a pair of time series (interest rate vs. house price index), we perform the self-consistent test in moving windows with size of $1$ to $4$ years. Periods that pass the tests are highlighted by shadows with colors encoding the $\rho$ values using a heat color scale. Red represents $\rho$ values close to zero (strong lead-lag signal) and blue corresponds to $\rho$ values close to one (weak lead-lag signal), following the same convention as in the signal strength map of Fig.~\ref{Fig:Signal Strength Map}. If a period possesses very strong lead-lag signals, there should exist many significant moving windows overlapping in its neighborhood. As each window is plotted by thin dot lines, the shadows of the strong signal period appear to be thick and dense.

\subsubsection{United Kingdom}

The general conclusion that can be extracted from Fig.~\ref{Fig:TOPS UK} is that the TOPS paths
$\langle x(t) \rangle$ are overwhelmingly negative, which indicates that interest rate changes are
lagging behind house price index changes over the period from 1991 to 2008. The lags are
however time dependent and are modulated by different macro-economic developments
that shocked the markets from 1991 to 2011, as explained below.
The overall information however is that the UK central bank has been mostly
``playing catch up'' with the real-estate markets, reacting to them rather than really
be on top of the game from 1991 to 2008. From 2008 to 2011, the lead-lag dependence
reverses, with $\langle x(t) \rangle$ being mainly positive, which corresponds to the
interest rates leading the real-estate markets. This confirms the change of regime
that occurred around 2008
with the financial crisis and great recession in which central banks have become
much more influential with their unconventional and very strong market interventions.

As shown in Fig.~\ref{Fig:TOPS UK}(\textsf{A}), the TOPS paths between the Bank of England's official Bank Rate (BR) and the Nationwide Building Society house price index (NBSHPI) exhibit a significant plateau from 1996 to 2000 with $\langle x(t) \rangle$ stabilizing at about $-5$. The TOPS paths during this period are highlighted by red shadows, to indicate that significant lead-lag structures are
qualified by the self-consistent tests based on Eq.~(\ref{Eq:Synchronized Self-Consistent}) and
strong lead-lag signals cannot be explained by random noise as proved by $\rho \approx 0$. Specifically,
our study suggests that, from 1996 to 2000, the monthly logarithmic returns of the Bank of England's Official Bank Rate $r_{\rm{BR}}$ and the monthly logarithmic returns of the Nationwide Building Society house price index $r_{\rm{NBSHPI}}$ are related by
the following lead-lag relationship
\begin{equation}
  R_{\rm{NBSHPI}}(t) = c + a*R_{\rm{BR}}(t-\langle x(t) \rangle) + \varepsilon(t),
\end{equation}
where $\langle x(t) \rangle \approx -5$ and the coefficient $a$ is strictly non-zero  during this period, indicating that $r_{\rm{NBSHPI}}$ was leading the development of $r_{\rm{BR}}$ with a time lag of about $5$ months. The steady rise of the house price index in this period suggests that housing market had stepped into its recovery phase. However, the recovery of the credit market did not go well as the housing market did. The failure of the Barings Bank impacted the credit market again, resulting in a further decline of interest rates in the years 1995 and 1996. The credit market recovered soon afterwards. Thus significant correlation signals within this period could be regarded as a phase of common recovery. We can also observe that the TOPS paths of BR-HHPI (\textsf{B}), BPLibor-NBSHPI (\textsf{C}), and BPLibor-HHPI (\textsf{D}) are all quite stable during this recovery period, with a time lag of half to one year. This reinforces the significant lead-lag signals between BR and NBSHPI. Together with the aftermath-adjusted phase, the period of late 1991 to 2000 could be identified as a common recovery phase, during which housing market was reviving steadily and leading the process while the credit market suffered shocks and lagged behind the housing market. On the other hand, it also draws the picture that the housing market played the role of assistor, leading the credit market recover from recessions after turbulence.

As illustrated in Fig.~\ref{Fig:TOPS UK}(\textsf{D}), the TOPS paths between the British pound LIBOR and the Halifax house price index present strong lead-lag signals from 1991 to 1995, which can be summarized by
\begin{equation}
  R_{\rm{HHPI}}(t) = c + a_{ }R_{\rm{BPLibor}}(t-\langle x(t) \rangle) + \varepsilon(t),
\label{Eq: TOPS HHPI BPLibor}
\end{equation}
where $\langle x(t) \rangle \approx -10$ and $a \neq 0$. According to the relationship, the Halifax house price index was leading the British pound LIBOR at a stable time lag about 10 months. This remarkable lead-lag structure shows that the development of the UK housing market as well as the declining and oscillating interest rates were still adjusting to the bubble bust of 1991. However, the decline of the house price index was slowing down and reached its bottom at the end of 1992. The housing market started to stabilize, while a new wave of decline of interest rates was triggered by the Sterling crisis at the end of 1992, making the interest rates drop again and reach their bottom towards the end of 1993. After the later shock of the Sterling crisis, the interest rates also went into an oscillation period associated with an adjustment to the moving house price index. Significant correlation signals during this period are detected as both of the two markets were recovering from the economy turbulence and about to settle down. Equation (\ref{Eq: TOPS HHPI BPLibor}) characterizes this lead-lag relationship caused by the second shock of the Sterling crisis on interest rates, which made the stabilization of interest rates come later than the house price index. The lead-lag structures during this period could be regarded as a phase of common adjustment to a series of shocks.

The signals associated with the recovery phases disappeared in 2000, when the dotcom bubble collapsed. During the period between 2000 and 2004, interest rates kept oscillating and declining into a very low level about $5\%$. In the meantime, the house price index experienced a period of rapid growth with a small plateau at the end of 2004, which probably was caused by the sudden depression of the credit market. The correlation between the two markets at this period was weak due to their entirely different developments, according to which the credit market was suffering from the dotcom bubble bust while the housing market was enjoying its golden time. Therefore, the TOPS paths during this period lost the significance of genuine lead-lag signals and kept falling toward the negative, which suggests that the growth of the house price indices were
developing and leading more and more the growth of interest rates.

The TOPS paths between all the short-to-medium term interest rates and the house price index exhibit remarkable lead-lag signals at the end of 2004, which can be summarized by
\begin{equation}
  R_{\rm{NBSHPI, HHPI}}(t) = c + a_{ }R_{\rm{BR, BPLibor}}(t-\langle x(t) \rangle) + \varepsilon(t),
\label{Eq: TOPS NBSHPI HHPI BR BPLibor}
\end{equation}
where $\langle x(t) \rangle$ started from the interval of $[-30, -20]$ and then went into a plateau of $[-3, -5]$ during 2007. During the period of 2004 to 2007, the house price index showed undoubtable signs of a bubble  \citep{Zhou-Sornette-2003a-PA,Zhou-Sornette-2006b-PA}. Recovered from the shock of the dotcom bubble, the interest rates also started to increase at the end of 2004. The lead-lag structure started at the end of 2004 with a large time lags, according to which the house price index was leading the interest rates by about 2 years, then went back rapidly toward zero and stabilized at a small time lags of about 3 to 4 months. This suggests that the interest rates were trying to ``catch up'' with the house price index, indicating that the Bank of England was rising its base rate, trying to control the wild growth of house price. This period could be identified as the phase when the Central Bank was attempting to cope with the overheating housing market.

Partially due to the subprime crisis of the United States in 2007, the UK housing market collapsed in 2008 and further participated to the triggering of the European debt crisis. Both the house price index and the interest rates dropped sharply in 2008. The TOPS paths in 2008 also run into a downward oscillation, indicating that the falling of the house price index was leading the falling of the interest rates in the 2008 turbulence. It identifies that the bust of the housing market bubble triggered the European debt crisis. Going through the year 2008,
$\langle x(t) \rangle$ in (\ref{Eq: TOPS NBSHPI HHPI BR BPLibor}) rebounded rapidly toward positive values. The interest rates turned to lead the house price index in 2009. It is a sign that the Bank of England was taking active monetary actions to slow down the collapsing housing market. The lead-lag structures returned to zero in 2010, suggesting that the evolution of interest rates and house price index after the economy turbulence tended to be synchronized, but with weakening correlation, as shown by the increasing $\rho$ value.

For the United Kingdom, the influence of the 10-year Government Bond Yield can be neglected since no significant lead-lag signals can be detected by its TOPS paths. It only has a theoretical impact on loans that are based on fixed rates. However, as proxies of adjustable rate mortgages, both the official bank rate and the 1-year British pound LIBOR are found to exhibit significant correlations with the house price index.

\subsubsection{United States}

\begin{figure}[ht]
\centering
  \includegraphics[width=14cm]{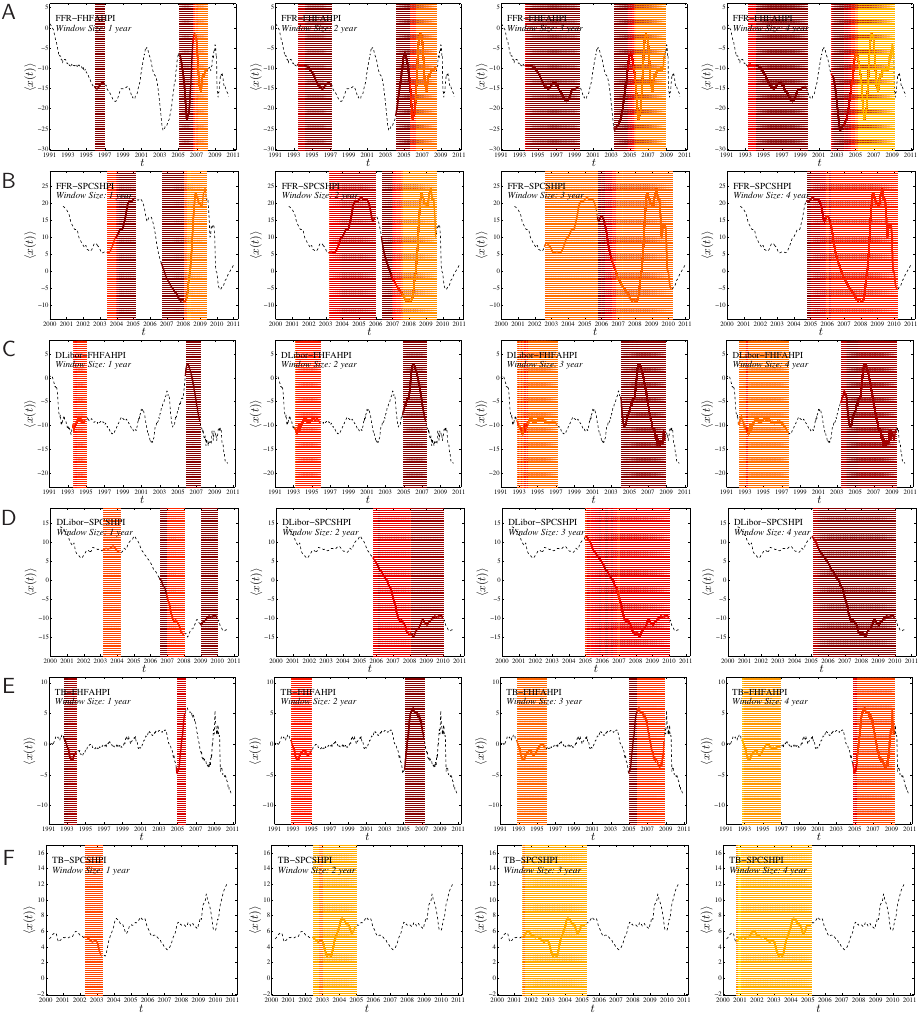}
  \caption{(color online) TOPS analysis of the  normalized returns $R(t)$ of monthly interest rate and of seasonal adjusted house price index of the US. The analysis is implemented at temperature $T=2$, using the distance definition $\epsilon_{-}$. Each row shows the result for a pair of ``interest rate vs. house price index'' series. The pairs FFR vs. FHFAHPI, FFR vs. SPCSHPI, DLibor vs. FHFAHPI, DLibor vs. SPCSHPI, TB vs. FHFAHPI, TB vs. SPCSHPI correspond respectively to rows \textsf{A},\textsf{B},\textsf{C},\textsf{D},\textsf{E},\textsf{F}. The self-consistent test is implemented within moving windows with sizes of $1$ to  $4$ years, corresponding to columns $1$ to $4$ respectively. Each black dash line is the resulting TOPS path $\langle x(t) \rangle$, which is chosen as the one with lowest free energy among $41 \times 41$ paths of different starting points ($t_X=i_1,t_Y=i_2$) and ending points ($t_X=N-i_1,t_Y=N-i_2$) for $i_1,i_2=0,1,2,\cdots,30$. The case $\langle x(t) \rangle > 0$ indicates that interest rate changes are preceding house price index changes at time $t$, and vice versa. Periods qualified by the self-consistent test are highlighted by shadow areas, whose color encode their corresponding $\rho$ values as in the signal strength map (Fig.~\ref{Fig:Signal Strength Map}, $T=2$). The $\rho$ values are calculated based on their estimated $\hat{a}$ and $\hat{f}$ according as explained in the text.}
  \label{Fig:TOPS US}
\end{figure}

Similarly to the UK case, Fig.~\ref{Fig:TOPS US} supports the conclusion that the TOPS paths $\langle x(t) \rangle$ are overwhelmingly negative until the crisis in 2006-2007, which indicates that interest rate changes were lagging behind house price index changes. The TOPS paths also suggests a catch up of the Federal Reserve still not being on top of the game during the crisis itself, as diagnosed by again the significant negative values of TOPS paths $\langle x(t) \rangle$ until 2008. Only later did the Federal Reserve and longer maturity rates lead the house price indices, confirming the occurrence of the transition to an era where the central bank is causally influencing the markets more than the reverse.

After the shock of the US Savings and Loans Crisis in 1991, the US interest rates went into a steady increasing phase from 1993 till 2000. In the meantime, the FHFA house price index went through a substantial growth phase. A very strong correlation signal in the period between 1993 and 2000 can be detected by the TOPS paths between the Federal Fund Rate and the FHFA house price index in Fig.~\ref{Fig:TOPS US}(\textsf{A}), which is well captured by the following relationship
\begin{equation}
r_{\rm{FHFA}}(t) = c + a_{ }r_{\rm{FFR}}(t-\langle x(t) \rangle) + \varepsilon(t),
\end{equation}
where $\langle x(t) \rangle$ started from about $-10$ and had a slight tendency to become more negative. This
suggests that the house price index was leading the Federal Fund Rate by a time lag of about $1$ year and
was rising faster and faster compared with the Federal Fund Rate during this period. In Fig.~\ref{Fig:TOPS US}(\textsf{C}), a relatively weak lead-lag signal can be detected by the TOPS paths between the dollar LIBOR and the FHFA house price index during this period,
which is summarized by
\begin{equation}
r_{\rm{FHFA}}(t) = c + a_{ }r_{\rm{DLibor}}(t-\langle x(t) \rangle) + \varepsilon(t),
\end{equation}
where $\langle x(t) \rangle \approx -10$ and $\rho \approx [0.1, 0.15]$. This suggests that the dollar LIBOR's growth rate was lagging behind the house price index by a stable lag time of about 10 months. There is however a probability of $10\% \sim 15\%$ that this signal
could result from noise. An even weaker lead-lag signal also shows up in the TOPS paths between the Treasury Bond Yield and the FHFAHPI during this period, represented by
\begin{equation}
r_{\rm{FHFA}}(t) = c + a_{ }r_{\rm{TB}}(t-\langle x(t) \rangle) + \varepsilon(t),
\end{equation}
where $\langle x(t) \rangle \approx -2$ and $\rho \approx [0.3, 0.4]$. This suggests that the Treasury Bond Yield was lagging behind the house price index by a smaller lag time of about 2 months. However, there is a probability of $30\% \sim 40\%$ that this signal
could be explained by noise. All these results signal a common prosperity phase between 1993 and 2000, which was led by the housing market.

After the bust of the dotcom bubble in 2000 and the 9/11 terrorist attack in 2001, the interest rates of the US exhibited waves of rapid and massive drops during the period between 2000 and 2004. In contrast, the house price index rose with an even faster growth rate, but evidence of a bubble was weak until 2004  \citep{Zhou-Sornette-2003a-PA}. The TOPS paths in this period lost their lead-lag signals. It is interesting to note that the FFR was found at that time to follow the US stock market antibubble  \citep{Sornette-Zhou-2005-QF,Zhou-Sornette-2006-JMe}.

The interest rates reached their bottoms in late 2003 and an aggressive rebound occurred between 2004 and 2007. The housing market, in the meantime, entered a phase of faster-than-exponential growth, showing solid signs of a bubble  \citep{Zhou-Sornette-2006b-PA}. The TOPS paths during this period exhibit significant lead-lag signals, which verged quickly toward zero, especially for the dollar LIBOR and the Treasury Bond Yield, whose TOPS paths become positive in the year around 2005 and 2006. This remarkable development of the lead-lag structures indicates that the Federal Reserve was attempting to cool down the overheating housing market by increasing the interest rates. This is quite similar to the UK case.

The US housing market bubble started to burst in late 2006 and triggered the infamous subprime crisis. Consecutively, the house price index and the interest rates crashed
towards the end of 2007. The TOPS paths during this time period oscillate and turn into negative values, suggesting that the crash of the house price index led the drop of the interest rates. After the collapse, the TOPS paths rapidly rebound toward positive values, indicating that interest rates were trying to catch up with the falling of the house price index and control the collapse. The lead-lag structures between 2007 and 2009 could be interpreted as a phase during which the crash of the housing market led the crash of the
interest rates and, later on, monetary policy
took over the causal relationship, trying to alleviate the consequence of the
crisis and of the collapsing housing market.

The lead-lag structures between the S$\&$P Case-Shiller house price index and the interest rates are quite different from those between the Federal Housing Finance Agency house price index and the interest rates. No consistent conclusion can be made. As shown in Fig.~\ref{Fig:TimeSeries Initial Return}({\textsf{D}}), the trajectories of the S$\&$P Case-Shiller HPI and the FHFA HPI have several significant discrepancies. First, the S$\&$P Case-Shiller HPI has a shorter span from 2000 to 2011. Second, the S$\&$P Case-Shiller HPI climaxed in late 2006 as successfully predicted in \citet{Zhou-Sornette-2006b-PA}, while the FHFA HPI peaked in 2007. Third, their return series have different patterns in several periods. While the TOPS paths between the FHFA HPI and the interest rates have very similar shapes with respect to different window sizes and for different maturities of the interest rates, the TOPS paths in the S$\&$P Case-Shiller HPI case are only stable with regard to different window sizes, but not for different interest rates. We thus conclude that the results obtained for the FHFA HPI are reasonable and convincing, but not for the S$\&$P Case-Shiller HPI.

\section{Discussion and conclusions}
\label{Sec:Conclusion}

In this work, we improved the thermal optimal path method (TOP) and introduced the symmetric thermal optimal path method (TOPS), making this novel technique more consistent and stable. We also introduced two powerful statistical tests specifically adapted to the TOPS method. Based on these techniques, we investigated the dynamic lead-lag interdependence between the housing market and the monetary policy for the case of two countries, namely United Kingdom and United States. Our study enhances the existing literature by providing a time-dependent resolution, together with reliable statistical test results, concerning the analysis of correlations between the data sets. The resulting lead-lag structures reveal three common phases for both countries, and some of these correlations are in line with other recent findings.

From 1991 to 1999, one can observe a phase of common recovery and prosperity between the housing market and credit market. Turbulence of the credit markets happened occasionally and impacted the interest rates, while the influence on house price indices is negligible. This draws a picture in which the housing market had been steadily growing and its rise fueled the increase of interest rates  \citep{Bernanke-Gertler-Gilchrist-1994-RES}.

The lead-lag signals of the common recovery phase disappeared in 2000 due to the international dotcom bubble bust. A new phase is detected that is characterized by significant lead-lag structures, during which interest rates are trying to catch up with the growth of house price index, identifying that both countries' central banks are making effort to cope with their respective overheating housing market. This overheating phase for the United Kingdom spans from 2004 to 2008, while for United States it goes from 2004 to 2007.

The collapse of the housing market triggered the subprime crisis in the United States in 2007-2008, which spread to the United Kingdom in 2008. Lead-lag structures during the crisis demonstrate that the crash of the housing market preceded the collapse of interest rates. However, after the slump, the development of interest rates happen to lead the house price indices, confirming by this metric that the central banks have taken more aggressive monetary policies to interfere more forcefully with the housing market as well with other economic variables, including the stock markets \citep{Guo-Zhou-Cheng-Sornette-2011-PLoS1}. While this observation may seem just to recover a well-known fact, this exercise confirms the soundness of the TOPS methodology. The value of these analyses lies in painting a richer picture than just a single causal inter-dependence that the Federal Reserve interest rate policy causes or is caused by the behavior of other economic variables. In fact, we have presented convincing evidence that the causality has reversed several times during the last 25 years, in relation with different market regimes. In order to fully understand monetary policy and economic dynamics, the TOPS approach stresses the importance of accounting for changes of regimes, so that similar pieces of information or policies may have drastically different impacts and developments, conditional on the economic, financial and geopolitical conditions. This study reinforces the view that the hypothesis of statistical stationarity in economics is highly questionable. This has tremendous implications for econometric works and for the understanding of economic systems.

As a final remark, we would like to discuss briefly the use of the word ``causality'' in this paper. As is now clear, the proposed TOPS method is designed to unveil the underlying dynamics of the lead-lag structure between two time series. It is a variation of a correlation type measure and thus does not imply that the leading process is actually causing the lagging process \citep{Sornette-Zhou-2005-QF}. The term ``causality'' applies to much more restrictive cases, illustrated for instance by Granger causality \citep{Ashley-Granger-Schmalensee-1980-Em,Geweke-1984,Chen-Rangarajan-Feng-Ding-2004-PLA} or to the nonlinear state space reconstruction of causation \citep{Sugihara-May-Ye-Hsieh-Deyle-Fogarty-Munch-2012-Science}. Nevertheless, it is well recognized that monetary policies and real-estate markets mutually influence each other. Thus, finding a statistical significant lead-lag relationship between monetary policies and real-estate markets, where one of them precedes the other one, is suggestive (but not a proof) of a  ``causality'' between these two time series. Our reference to causal influence should be read with this caveat in mind.

\section*{Acknowledgement}

This work started initially with the Master Thesis of Jonas Nikolaus Debatin performed at ETH Zurich in 2011 and evolved into completely novel methods and results. We are grateful to the referees for their constructive suggestions. Possible remaining issues are our responsibility. This work was supported in part by the National Natural Science Foundation of China (71131007, 71501072 and 71532009).

%


\end{document}